\documentclass[12pt,article]{JHEP3}

\usepackage{amscd,amsmath,amssymb,amsfonts,xspace,mathrsfs}

\usepackage{epsf}

\def\lfig#1#2#3#4#5{
\begin{figure}
\refstepcounter{figure}
\label{#4}
\addtocounter{figure}{-1}
\epsfxsize=#3
\centerline{\epsfbox{#2}}
\vspace{#5}
{\bf \caption{{\rm #1}}}
\end{figure}
}

\def\twofig#1#2#3#4#5#6#7{
\begin{figure}
\refstepcounter{figure}
\label{#5}
\addtocounter{figure}{-1}
\centerline{\epsfxsize=#4\epsfbox{#2}
\hspace{#6}
\epsfxsize=#4\epsfbox{#3}}
\vspace{#7}
{\bf \caption{{\rm #1}}}
\end{figure}
}

\def\twofigdif#1#2#3#4#5#6#7#8{
\begin{figure}
\refstepcounter{figure}
\label{#5}
\addtocounter{figure}{-1}
\centerline{\epsfxsize=#4\epsfbox{#2}
\hspace{#6}
\epsfxsize=#8\epsfbox{#3}}
\vspace{#7}
{\bf \caption{{\rm #1}}}
\end{figure}
}

\def\threefigmod#1#2#3#4#5#6#7#8{
\begin{figure}
\refstepcounter{figure}
\label{#7}
\addtocounter{figure}{-1}
\vspace{2cm}
\epsfxsize=#5\epsfbox{#2}

\vspace{-11cm}\hspace{9.5cm}
\epsfxsize=#6\epsfbox{#3}

\hspace{9.5cm}
\epsfxsize=#6\epsfbox{#4}
\vspace{#8}
{\bf \caption{{\rm #1}}}
\end{figure}
}

\def\varpi{t}

\def\det{\,{\rm det}\, }

\def\Im{\,{\rm Im}\,}
\def\Re{\,{\rm Re}\,}

\def\({\left(}
\def\){\right)}
\def\[{\left[}
\def\]{\right]}
\def\<{\left\langle}
\def\>{\right\rangle}
\def\hf{{1\over 2}}

\newcommand{\CP}{\IC P^1}

\newcommand{\bfk}{{\boldsymbol k}}

\renewcommand{\d}{\mathrm{d}}
\newcommand{\de}{\mathrm{d}}

\newcommand{\I}{\mathrm{i}}

\newcommand{\cL}{\mathcal{L}}

\newcommand{\p}{\partial}

\newcommand{\cV}{\mathcal{V}}

\newcommand{\cK}{\mathcal{K}}
\newcommand{\cM}{\mathcal{M}}

\newcommand{\cN}{\mathcal{N}}
\newcommand{\cE}{\mathcal{E}}
\newcommand{\cX}{\mathcal{X}}

\newcommand{\cR}{\mathcal{R}}

\newcommand{\cJ}{\mathcal{J}}
\newcommand{\cZ}{\mathcal{Z}}

\newcommand{\cA}{\mathcal{A}}
\newcommand{\cB}{\mathcal{B}}

\newcommand{\cY}{\mathcal{Y}}

\def\mE{\mathscr{E}}

\def\mQ{\mathscr{Q}}

\def\mN{\mathscr{N}}

\DeclareSymbolFont{AMSa}{U}{msa}{m}{n}
\DeclareSymbolFont{AMSb}{U}{msb}{m}{n}
\DeclareMathSymbol{\fieldR}{\mathalpha}{AMSb}{"52}

\newcommand{\N}{{\mathcal N}}

\newcommand{\nn}{\nonumber}

\newcommand{\IR}{\mathbb{R}}
\newcommand{\IC}{\mathbb{C}}
\newcommand{\IZ}{\mathbb{Z}}

\newcommand{\tzeta}{\tilde\zeta}

\newcommand{\thh}{\tilde h}
\newcommand{\tlk}{\tilde k}

\def\bea{\begin{eqnarray}}
\def\eea{\end{eqnarray}}
\def\be{\begin{equation}}
\def\ee{\end{equation}}
\def\ba{\begin{align}}
\def\ea{\end{align}}
\def\bse{\begin{subequations}}
\def\ese{\end{subequations}}

\def\bX{\bar X}
\def\bF{\bar F}
\def\bY{\bar Y}

\def\bZ{\bar Z}
\def\bE{\bar E}

\def\ba{\bar a}

\def\bi{\bar \imath}
\def\bj{\bar \jmath}
\def\bk{\bar k}

\def\bn{\bar n}

\def\bz{\bar z}

\def\hN{\hat N}

\def\ci#1{c^{[#1]}}
\def\cij#1{c^{[#1]}}

\newcommand{\Li}{{\rm Li}}

\def\ellg#1{\ell_{#1}}

\def\nk{n_{k}^{(0)}}

\def\Zg#1{Z_{\gamma_{#1}}}
\def\bZg#1{\bar Z_{\gamma_{#1}}}

\def\hng#1{\Omega_{\gamma_{#1}}}
\def\Om#1{\Omega_{#1}}

\def\Ilg{\cJ^{(1)}}
\def\Ilog#1{\cJ^{(1,#1)}}
\def\Irt{\cJ^{(2)}}
\def\Irat#1{\cJ^{(2,#1)}}

\def\Igp{\Ilog{+}_{\gamma}}
\def\Igm{\Ilog{-}_{\gamma}}
\def\Igpm{\Ilog{\pm}_{\gamma}}

\def\Igg#1{\Ilg_{\gamma_{#1}}}

\def\rIg{\Irt_{\gamma}}
\def\rIgp{\Irat{+}_{\gamma}}
\def\rIgm{\Irat{-}_{\gamma}}
\def\rIgpm{\Irat{\pm}_{\gamma}}

\def\Ingam#1#2{\cJ^{(#1)}_{#2}}
\def\Insgam#1#2#3{\cJ^{(#1,#2)}_{#3}}

\def\cij#1{c}
\def\ci#1{c}

\def\Mb{\mathbf{M}}
\def\mb{\mathbf{m}}

\def\vl{v}
\def\bvl{\bar \vl}

\def\Min{M}

\def\Uin{\mathbf{U}}

\def\cVs{\cV_{(\sigma)}}

\def\cMsk{\cM_{\rm sk}}
\def\CY{\mathfrak{Y}}

\def\Fcl{F^{\rm cl}}
\def\bFcl{\bF^{\rm cl}}

\def\gp{\gamma}

\def\Mbi#1{\Mb^{(#1)}}

\def\Va{V^{(\varphi)}}
\def\Ninst{N_{\rm inst}}

\preprint{L2C:16-094 \\ IPMU16-0091}

\title{
Non-perturbative scalar potential inspired by\\ type IIA strings on rigid CY
}

\author{Sergei Alexandrov$^{a}$, Sergei V. Ketov$^{b,c,d}$ and Yuki Wakimoto$^{b}$
\\
$^a$ {\it
Laboratoire Charles Coulomb (L2C), UMR 5221, CNRS-Universit\'e de
Montpellier, F-34095, Montpellier, France}\\

$^b$ {\it Department of Physics, Tokyo Metropolitan University,
1-1 Minami-ohsawa, Hachioji-shi, Tokyo 192-0397, Japan}\\

$^c$ {\it Kavli Institute for the Physics and Mathematics of the Universe (IPMU),
The University of Tokyo, Chiba 277-8568, Japan} \\

$^d$ {\it Institute of Physics and Technology, Tomsk Polytechnic University,
30 Lenin Ave., Tomsk 634050, Russian Federation
} \\

\vspace*{2mm} {\tt e-mail:
\email{salexand@univ-montp2.fr},
\email{ketov@tmu.ac.jp},
\email{wakimoto-yuki@ed.tmu.ac.jp}
}

\vspace*{-3mm}

}

\abstract{Motivated by a class of flux compactifications of type IIA strings on rigid Calabi-Yau manifolds,
preserving $N=2$ local supersymmetry in four dimensions, we derive a non-perturbative potential of all scalar fields
from the exact D-instanton corrected metric on the hypermultiplet moduli space.
Applying this potential to moduli stabilization, we find a discrete set of exact vacua for axions.
At these critical points, the stability problem is decoupled into two subspaces
spanned by the axions and the other fields (dilaton and K\"ahler moduli), respectively.
Whereas the stability of the axions is easily achieved,
numerical analysis shows instabilities in the second subspace.
}

\begin{document}

\section{Introduction}
\label{sec-intro}

One of the outstanding issues in string theory is the problem of finding realistic string compactifications
and connecting them to cosmological observations.
It requires several steps such as (i) choosing an appropriate setup for moduli stabilization, (ii)
obtaining a meta-stable vacuum with a positive cosmological constant, and (iii) producing
an inflationary model. Each of these steps is highly non-trivial and has its own obstructions.
Despite of many years of research and the extensive literature on the subject,
meta-stable de Sitter (dS) vacua still appear to be very difficult to get in string theory.
There are no robust predictions about inflation, and no nice inflationary model from string theory was found yet.
And both, dS vacua and inflation, are usually obtained in string theory at the price of adding effects which can spoil
moduli stabilization (see \cite{Baumann:2014nda} for a recent review).

Furthermore, most of the scenarios in string theory cannot be considered as those
derived from the first principles,  because of at least one of the following reasons:
\begin{itemize}
\item
the lack of precise knowledge about quantum corrections,
\item
splitting the procedure of moduli stabilization into several steps which may result in ignorance
of tachyonic directions spoiling meta-stability,
\item
the necessity to introduce additional uplifting mechanisms,
\item
disregarding back reaction effects.
\end{itemize}
The first of these issues is particularly important.
While it is possible to stabilize all moduli at the classical level \cite{DeWolfe:2005uu},
several no-go theorems forbid dS vacua in such simplest supergravity compactifications \cite{Maldacena:2000mw,Ivanov:2000fg}.
To avoid them, it is necessary to include either quantum corrections, both perturbative and non-perturbative,
or non-geometric fluxes (see, for instance,
\cite{Kachru:2003aw,Balasubramanian:2005zx,Conlon:2005ki,Westphal:2006tn,deCarlos:2009qm,Louis:2012nb,Danielsson:2012by,Blaback:2013qza,Hassler:2014mla}).

The significance of explicit examples of truly ``quantum" calculations in string theory
goes well beyond the problem of the cosmological constant.
It is just about the string theory based computations of quantum gravity corrections that are usually
put ``out of brackets" in modern phenomenologically based theoretical cosmology.
Taking into account non-perturbative corrections is necessary to stabilise all moduli,
provide resolution of unphysical singularities in moduli spaces, and ensure string dualities.
The very possibility of explicit (or exact)
non-perturbative calculations is highly non-trivial in string theory, and the known examples are very rare.

One example, where such calculations have become possible, is the case of type II string compactifications
on Calabi-Yau (CY) threefolds. In this case the low energy effective action (LEEA)
in four dimensions preserves $N=2$ local supersymmetry (8 supercharges) and is completely determined
by the geometry of its moduli space spanned by the scalar fields of $N=2$ vector and hypermultiplets.
While the vector multiplet moduli space was described in full detail using mirror symmetry long ago
(see, e.g., \cite{VanProeyen:1995sw} for a review), understanding of the quantum corrected hypermultiplet moduli
space was very limited until recently. The advance of twistorial techniques drastically changed the situation
and allowed us to get an {\it exact} description of the most of quantum effects ---
at present, amongst all quantum corrections, only the so-called NS5-brane instantons remain out of control
(see \cite{Alexandrov:2011va,Alexandrov:2013yva} and references therein).

Thus, it is natural to apply these exact results in a more general context of moduli stabilization.
Of course, this requires
extending them beyond the class of compactifications where they were initially derived.
In particular, the phenomenologically interesting compactifications include fluxes, localized sources
such as D-branes and orientifold planes, and preserve only $N=1$ local supersymmetry (4 supercharges) in four dimensions.
However, at present, quantum corrections are beyond control in such cases.

On the other hand, it is possible to generate a non-trivial scalar potential for moduli stabilization in a unique way
already in $N=2$ supergravity. 
This can be achieved by adding NS- and RR-fluxes leading to the gauging of some of the isometries
of the moduli space of the original fluxless compactification. In fact, the integrated Bianchi identities give rise
to certain tadpole cancellation conditions, which in the presence of fluxes
generically can be satisfied only by adding orientifolds reducing supersymmetry to $N=1$ \cite{Giddings:2001yu}.
However, in type IIA string theory it is possible to choose such fluxes that the tadpole cancellation condition holds automatically.

This motivates us to consider $N=2$ gauge supergravity, which results from
the type IIA CY compactifications with the NS $H$-fluxes and the RR $F_4$- and $F_6$-fluxes
provided one ignores their back reaction.
Such setup was already studied in \cite{Kachru:2004jr}. We go beyond the earlier studies, and
compute the quantum corrected scalar potential in the gauged supergravity including the non-perturbative terms,
which come from the instanton corrections to the geometry of the moduli space known exactly in the absence of fluxes.
The idea beyond this computation is that the preserved $N=2$ supersymmetry protects the quantum corrections 
so that the exact non-perturbative potential, where the back reaction effects are taken into account, 
should not differ too much from the one obtained here.

In this paper we restrict ourselves to the case of a {\it rigid} CY threefold $\CY$.
Such manifold has the vanishing Hodge number $h^{2,1}(\CY)=0$, so that the LEEA
is described by $N=2$ supergravity interacting with a single hypermultiplet, called the {\it universal hypermultiplet} (UH),
and some number $h^{1,1}(\CY)> 0$ of vector multiplets. This leads to various simplifications,
such as the absence of complex structure moduli, which allow
to make our analysis very explicit. Actually, one of our original motivations was to find
a setup for flux compactifications which takes into account quantum corrections and,
at the same time, can be treated as explicitly as possible.

It should be mentioned that several attempts to take into account instanton corrections in compactifications
on rigid CY  already appeared in the literature, most notably, in \cite{Davidse:2005ef}.
However, the analysis of \cite{Davidse:2005ef} did not include contributions of vector multiplets
and, as it turned out later, was based on a misleading ansatz for D-instantons.
In contrast, we consider here the full scalar potential including all moduli.
Moreover, we do not assume that there exists a hierarchy allowing us to perform moduli stabilization
in a step-by-step procedure, but analyze all equations on critical points on the same footing.

One of our results is a simple condition on the flux parameters (see \eqref{relflux})
which allows us to find a set of {\it exact} solutions to the quantum corrected equations for all axion fields,
i.e. the periods of the $B$-field and the RR 3-form potential along 2 and 3-cycles of $\CY$, respectively.
The role of the worldsheet and D-instanton corrections for the existence of these solutions is pivotal.

Unfortunately, the equations we get on the remaining scalars, namely, dilaton and K\"ahler moduli,
are too complicated to be treated in full generality.
Therefore, in the beginning we restrict our attention to the perturbative approximation
where all instanton contributions are neglected, but perturbative $\alpha'$ and $g_s$-corrections,
controlled by the Euler characteristic of $\CY$, are retained. We obtain bounds on the values of the dilaton and the CY volume,
which admit the existence of critical points. In particular, we find that this class of compactifications
does not allow critical points with both large volume and small string coupling, i.e. in the only region
where all quantum corrections can be neglected.
This can be contrasted with the result of \cite{DeWolfe:2005uu} that a more general choice of fluxes
provides the moduli stabilization at classical level, but this choice must be supplemented by an orientifold projection
to satisfy the tadpole cancellation condition mentioned above and leads to AdS vacua.

To further analyze the critical points, we first restrict ourselves to the case with one K\"ahler modulus,
i.e. to a CY with $h^{1,1}=1$. Since up to now no CY was found with such Hodge numbers, this case should only be 
viewed as a model convenient to test the moduli stabilization, but not having a string theory realization.
In this special case we find two critical points,
which both lead to a positive potential, but both turn out to be unstable.
Then we turn to the general case, where we directly address the problem of stability of critical points,
without trying to find them explicitly. To this end, we analyse the matrix of the second derivatives
and show that it cannot be positive definite, which means that there are no meta-stable vacua.
Thus, in the perturbative approximation, these simple models cannot provide stabilization of all moduli.

Finally, we attempt to take into account the contributions of worldsheet and D-instantons
in the simplest case of $h^{1,1}=1$. As before, we perform a numerical analysis of the second derivative
matrix, which shows us again that in the physical region the matrix is never positive definite on mass shell.
This result appears to be extremely non-trivial, given a very complicated analytical form of the second derivatives.
The effect of instantons on the perturbative analysis for $h^{1,1}>1$ will be investigated elsewhere.

The paper is organized as follows. In the next section we review some basic information about
CY string compactifications, their moduli spaces, the effect of fluxes, and provide
a formula for the scalar potential induced by the gauging in $N=2$ supergravity.
We also compute this potential explicitly, including
perturbative and non-perturbative quantum corrections, in the gauged supergravity inspired 
by the class of compactifications we concentrate on.
In section \ref{sec-modstab} we discuss equations on critical points and find a solution for all axion fields.
In section \ref{sec-pert} we study the perturbative approximation. First, we derive general bounds
on critical points, then analyze in detail the case with one K\"ahler modulus, and finally perform
a stability analysis in a generic case with arbitrary number of moduli.
In section \ref{sec-inst} we present the results of our numerical analysis
of the one-modulus case in the presence of instantons. Section \ref{sec-concl}
is devoted to a discussion of our results. Several appendices contain details
about special and quaternionic geometries, the metrics on $N=2$ vector and hypermultiplet
moduli spaces, and our stability analysis of critical points.

\section{Scalar potential from gauging}
\label{sec-potential}

\subsection{$N=2$ gauged supergravity and its scalar potential}
\label{subsec-SUGRA}

The four-dimensional LEEA of type II strings compactified on a Calabi-Yau threefold $\CY$
is given by $N=2$ supergravity coupled to $N=2$ vector and hypermultiplets.
In the two-derivative approximation, where one ignores the higher curvature terms appearing as $\alpha'$-corrections,
the bosonic part of the action comprises only kinetic terms for the metric, vector and scalar fields
arising after compactification. The couplings of these kinetic terms are, however, non-trivial, being
restricted by $N=2$ supersymmetry in terms of the metrics on the vector and hypermultiplet moduli spaces,
$\cM_V$ and $\cM_H$, parametrized by the scalars of the corresponding multiplets.
Furthermore, $N=2$ supersymmetry restricts $\cM_V$ to be a special K\"ahler manifold, with a K\"ahler potential $\cK(z^i,\bz^{\bi})$
(with $i=1,\dots, h^{1,1}$ in type IIA) determined by a holomorphic prepotential $F(X^I)$
(with $I=(0,i)=0,\dots ,h^{1,1}$ and $z^i=X^i/X^0$),
a homogeneous function of degree 2. Similarly, $\cM_H$ must be a quaternion-K\"ahler (QK) manifold of dimension
$4(h^{2,1}+1)$ \cite{Bagger:1983tt}.
We denote the metrics on the two moduli spaces by $\cK_{i\bj}$ and $g_{uv}$, respectively.

The resulting theory is, however, not appropriate from the phenomenological point of view since
it does not have a scalar potential, so that all moduli remain unspecified.
This gives rise to the problem of {\it moduli stabilization}, i.e. generating a potential for the moduli
with a local minimum and no flat directions.
Local $N=2$ supersymmetry does allow a non-trivial scalar potential, but this requires to consider
$N=2$ {\it gauged} supergravity.
The latter can be constructed from the usual ungauged supergravity when the moduli space $\cM_V\times \cM_H$ has some isometries,
which are to be gauged with respect to the vector fields $A^I$ comprising,
besides those of vector multiplets, the gravi-photon $A^0$ of the gravitational multiplet.
Physically, this means that the scalar fields affected by the isometries acquire charges
under the vector fields used in the gauging.
The charges are proportional to the components of the Killing vectors $k_\alpha$ corresponding to the gauged isometries.
In general, the gauge group must be a subgroup of the isometry group, but
in this paper we deal only with abelian gaugings of isometries of the hypermultiplet moduli space $\cM_H$.
Then the charges are characterized by the vectors
$\bfk_I=\Theta_I^\alpha k_\alpha\in T\cM_H$ where $\Theta_I^\alpha$ is known as the {\it embedding tensor}.

It is remarkable that in $N=2$ gauged supergravity the geometry of the moduli space together with the charge vectors
{\it completely} fix the scalar potential. Explicitly, it is given by
\cite{D'Auria:1990fj,Andrianopoli:1996cm,deWit:2001bk}\footnote{Our conventions
and normalizations are explained in Appendix \ref{ap-norm}.
Note that the potential appears in the literature in the two possible forms, which
are both given in \eqref{scpot-generic} and are simply related by Eq. \eqref{relcND}.
In the presence of non-abelian gaugings the potential acquires additional terms which we, however, omit.}
\bea
V &=&
4 e^\cK \bfk^u_I \bfk^v_J g_{uv} X^I \bX^J + e^\cK \(\cK^{i\bj}D_i X^I D_{\bj}\bX^J-3 X^I \bX^J\)\(\vec\mu_I\cdot \vec \mu_J\),
\label{scpot-gen}
\eea
where $D_i X^I=(\p_i+\p_i \cK)X^I$
and $\vec\mu_I$ is the triplet of moment maps which quaternionic geometry of $\cM_H$ assigns to each isometry $\bfk_I$ \cite{MR872143}.
This result gives us an opportunity to search for the potentials ensuring moduli stabilization,
using the geometric data from the ungauged theory as an input.
In particular, here we employ the exact results about the non-perturbative description of $\cM_V$ and $\cM_H$ in
type II CY compactifications, described below in subsection \ref{subsec-qcor},
to infer the impact of quantum corrections on the potential \eqref{scpot-gen} and stabilization of moduli.

\subsection{Flux compactifications}
\label{subsec-flux}

In string theory, $N=2$ gauge supergravity can be obtained by adding closed string fluxes to a CY compactification
(see \cite{Grana:2005jc} for a review).
In fact, fluxes back react on the background geometry so that the simple direct product $M_4\times \CY$ is not a solution of the (classical)
equations of motion anymore. To get a solution, one has to add a warp factor and to consider internal manifolds with torsion
\cite{Strominger:1986uh,Polchinski:1995sm,Michelson:1996pn}. Although such backgrounds are nicely described in the framework
of generalized geometry \cite{Hitchin:2004ut}, the corresponding effective actions are poorly understood.
Due to this reason, we accept the common strategy (see, for instance,
\cite{Louis:2002ny,Giryavets:2003vd,Kachru:2004jr,DeWolfe:2005uu})
and ignore the back reaction, assuming that the compactification manifold is still a Calabi-Yau.\footnote{It should be mentioned
that in the type IIA theory under consideration in this paper, it is the less justified assumption than in type IIB.
In the latter case, some choices of fluxes allow the vacua where the internal
manifold is a {\it conformal} Calabi-Yau space, which is not too much different from the usual Calabi-Yau manifolds.
In contrast, in the type IIA case the equations of motion require the compactification manifold
to be either non-K\"ahler, or even non-complex.}

The LEEA for flux compactifications on CY was found in \cite{Louis:2002ny},
and was shown to perfectly fit the framework of $N=2$ gauged supergravity.\footnote{More precisely,
in the presence of the so-called magnetic fluxes, it should be generalized to incorporate massive tensors.
In the absence of fluxes, these tensor fields are massless and can be dualized to
the scalars contributing to the hypermultiplet moduli space. After receiving a mass,
they are rather dual to massive vector fields.\label{foot-magnetic}}
In particular, given the LEEA, one can read off the embedding tensor $\Theta_I^\alpha$ providing
a map between the fluxes and the gauged isometries. Let us briefly review these results.

First, we recall the field content of the moduli spaces. In type IIA, the vector multiplet moduli space
$\cM_V$ describes the complexified K\"ahler moduli of $\CY$ parametrizing deformations of the K\"ahler structure and
the periods of the $B$-field along two-dimensional cycles, $z^i=b^i+\I t^i$.
The hypermultiplet moduli space $\cM_H$ consists of
\begin{itemize}
\item
$u^a$ --- complex structure moduli of $\CY$ ($a=1,\dots,h^{2,1}$),
\item
$\zeta^\Lambda,\tzeta_\Lambda$ --- RR-scalars given by periods of the RR 3-form potential along three-dimensional cycles of $\CY$
($\Lambda=(0,a)=0,\dots,h^{2,1}$),
\item
$\sigma$ --- NS-axion, dual to the 2-form $B$-field in four dimensions,
\item
$\phi$ --- dilaton, determining the value of the four-dimensional string coupling, $g_s^{-2}=e^{\phi}\equiv r$.
\end{itemize}

The Kaluza-Klein reduction from ten dimensions, performed in \cite{Louis:2002ny}, leads to the classical metrics on $\cM_V$
and $\cM_H$. The former is the special K\"ahler metric $\cK_{i\bj}$ given by the derivatives of the K\"ahler potential
\be
\cK=-\log\[ \I\(\bX^I \Fcl_I-X^I\bFcl_I\)\],
\ee
where $\Fcl_I=\p_{X^I}\Fcl$ are the derivatives of the classical holomorphic prepotential
\be
\Fcl(X)=-\kappa_{ijk}\, \frac{X^iX^j X^k}{6X^0},
\label{Fcl}
\ee
which is determined by the triple intersection numbers $\kappa_{ijk}$ of $\CY$.
The hypermultiplet metric is given by the so-called {\it c-map} \cite{Cecotti:1989qn} which produces
a QK metric out of another holomorphic prepotential characterizing the complex structure moduli.
We omit its explicit expression, but mention the crucial fact that it carries
a Heisenberg group of continuous isometries acting by shifts on the RR-scalars and the NS-axion.
The corresponding Killing vectors are
\be
\label{heis0}
k^\Lambda=\p_{\tzeta_\Lambda}-\zeta^\Lambda\p_\sigma,
\qquad
\tlk_\Lambda=\p_{\zeta^\Lambda}+\tzeta_\Lambda\p_\sigma,
\qquad
k_\sigma=2\p_\sigma.
\ee
It is these isometries that are gauged by adding fluxes.

In general, type IIA strings on CY admit NS-fluxes incorporated by the following field strength of the $B$-field:
\be
H^{\rm flux}_3=h^\Lambda\tilde\alpha_\Lambda-\thh_\Lambda\alpha^\Lambda,
\ee
where $(\alpha^\Lambda,\tilde\alpha_\Lambda)$ is a symplectic basis of harmonic 3-forms,
and RR-fluxes given by the 2- and 4-form field strengths
\be
F^{\rm flux}_2=-m^i\tilde\omega_i,
\qquad
F^{\rm flux}_4=e_i\omega^i,
\ee
where $\tilde\omega_i$ and $\omega^i$ are bases of $H^2(\CY)$ and $H^4(\CY)$, respectively.
Besides, there are two additional parameters, $m^0$ and $e_0$.
The first one is Romans mass which gives a consistent deformation of ten-dimensional type IIA supergravity \cite{Romans:1985tz},
and the second one is a constant arising after dualization of the 3-form RR potential \cite{Louis:2002ny}.
They can be viewed as the fluxes $F^{\rm flux}_0$ and $F^{\rm flux}_6$, and also lead to a gauging in the effective action.

Although the effective action was found in \cite{Louis:2002ny} in the presence of all these flux parameters,
we set the ``magnetic" fluxes $m^I$ to zero in what follows. The reason is twofold. First,
this allows to avoid complications with the simultaneous appearance of electric and magnetic charges of the NS-axion
as well as massive vector fields (see footnote \ref{foot-magnetic}).
Second, the vanishing of Romans mass $m^0$ allows to avoid adding orientifold planes, otherwise,
needed to satisfy the D6-brane tadpole cancellation condition \cite{Kachru:2004jr}.
This also allows us to keep $N=2$ supersymmetry unbroken, which partially justifies
our use of the results obtained for fluxless CY compactifications.

With this restriction, the gauging induced by the fluxes is
characterized by the following charges \cite{Louis:2002ny}:
\be
\bfk_0=h^\Lambda \tlk_\Lambda+\thh_\Lambda k^\Lambda+e_0k_\sigma,
\qquad
\bfk_i=e_i k_\sigma,
\label{ch-gauge}
\ee
written down here as linear combinations of the Killing vectors \eqref{heis0}.

\subsection{Quantum corrections}
\label{subsec-qcor}

The scalar potential obtained in \cite{Louis:2002ny} was found by the Kaluza-Klein reduction and,
therefore, resulted from gauging of the isometries of the {\it classical} moduli space. However,
both $\cM_V$ and $\cM_H$ are known to receive {\it quantum} corrections.
Unfortunately, one has a very limited understanding of the impact of fluxes on these corrections.
On the other hand, for fluxless CY compactifications the situation is much better, as we now describe.

We have full control over the metric on $\cM_V$: it receives the $\alpha'$-corrections which
are all captured by a modification of the holomorphic prepotential \eqref{Fcl} \cite{Candelas:1990rm,Hosono:1993qy}
\be
F(X)=\Fcl(X)+\chi_\CY\,\frac{\I\zeta(3)(X^0)^2}{16\pi^3}
-\frac{\I(X^0)^2}{8\pi^3}\sum_{k_i\gamma^i\in H_2^+(\CY)}\nk \Li_3\(e^{2\pi\I k_iX^i/X^0}\),
\label{Ffull}
\ee
where $\chi_\CY=2(h^{1,1}-h^{2,1})$ is Euler characteristic of CY, $\nk$ are the genus-zero Gopakumar-Vafa invariants,
and the sum goes over the effective homology classes, i.e. $k_i\ge 0$ for all $i$, with not all of them vanishing simultaneously.
The two additional terms correspond to a perturbative correction and a contribution of worldsheet instantons, respectively.

As regards $\cM_H$, though its complete non-perturbative description is still beyond reach, a significant progress
in this direction was recently achieved by using twistorial methods (see \cite{Alexandrov:2011va,Alexandrov:2013yva} for reviews).
In contrast to $\cM_V$, the hypermultiplet metric is exact in
$\alpha'$, but receives $g_s$-corrections. At the perturbative level, it is known explicitly \cite{Alexandrov:2007ec}
and is given by a one-parameter deformation of the classical c-map metric,
whose  deformation parameter is controlled by $\chi_\CY$ \cite{Antoniadis:1997eg,Antoniadis:2003sw,RoblesLlana:2006ez}.
At the non-perturbative level, the metric gets the instanton contributions coming from D2-branes wrapping 3-cycles
(and, hence, parametrized by a charge $\gamma=(p^\Lambda, q_\Lambda)$) and NS5-branes wrapping the whole CY.
The D-instantons were incorporated to all orders using the twistor description of QK manifolds
\cite{RoblesLlana:2006is,Alexandrov:2008nk,Alexandrov:2008gh,Alexandrov:2009zh}, so that
only NS5-instanton contributions still remain unknown
(see, however, \cite{Alexandrov:2010ca,Alexandrov:2014mfa,Alexandrov:2014rca}
for a recent progress on the type IIB side). Though the twistor description is rather implicit via
encoding the metric into the holomorphic data on the twistor space of $\cM_H$,
in the case when only the D-instantons with ``mutually local charges" $\langle\gamma,\gamma'\rangle=0$\footnote{We use
the skew symmetric product defined by $\langle\gamma,\gamma'\rangle=q_\Lambda p'^\Lambda-q'_\Lambda p^\Lambda$.
The mutual locality is equivalent to the condition that there is a symplectic frame where all charges are purely electric, i.e. $p^\Lambda=0$.}
are taken into account, the metric was explicitly computed in \cite{Alexandrov:2014sya}.

Thus, it is natural to use these exact results for analyzing the scalar potential \eqref{scpot-gen}.
Of course, it would be naive to expect that they are not going to be affected by fluxes and, eventually, their back reaction via torsion,
and it is an open question whether in such situation one can trust the quantum corrections computed before the fluxes were switched on.
However, the presence of $N=2$ supersymmetry allows us to think that the back reaction effects should not be too strong.
Indeed, most of the results mentioned above were obtained by using only requirements of supersymmetry and
a few discrete symmetries expected to survive at the non-perturbative level. Besides, this expectation
is supported by the recent results about perturbative $\alpha'$ and $g_s$-corrections
for compactifications on manifolds with the $SU(3)$ structure \cite{Grana:2014vva}.
In the worst case, if our expectation does turn out to be wrong, the gauged supergravity obtained in 
this approximation and studied in this paper should only be considered as inspired by string theory.

It should be noticed that instanton corrections break the continuous isometries
of the classical hypermultiplet moduli space: a D-instanton of charge
$\gamma$ comes with a factor $e^{2\pi\I(p^\Lambda\tzeta_\Lambda-q_\Lambda\zeta^\Lambda)}$
and, therefore, breaks a linear combination of $k^\Lambda$ and $\tlk_\Lambda$,
whereas NS-brane instantons break all isometries of \eqref{heis0}.
This raises the question, how such instantons can be consistent with the gauging induced by fluxes,
since the latter can be only performed in the presence of continuous isometries?
This problem was solved in \cite{KashaniPoor:2005si} where it was shown that
fluxes protect from the instanton corrections precisely those isometries that are to be gauged.
Applying this result to type IIA string theory on CY with $H_3$, $F_4$ and $F_6$ fluxes, one concludes from \eqref{ch-gauge}
that it excludes NS5-instantons and allows only D-instantons with charges satisfying $h^\Lambda q_\Lambda-\thh_\Lambda p^\Lambda=0$.

\subsection{Scalar potential from fluxes on rigid CY}
\label{subsec-fluxrigid}

In this paper we restrict our attention to the flux compactifications on a {\it rigid} Calabi-Yau manifold, i.e. when $\CY$
has vanishing $h^{2,1}$ and thus does not have complex structure deformations.
As a result, the capital Greek indices $\Lambda,\Sigma,\dots$
take only one value and, therefore, can be safely dropped.

In the case of rigid CY, $\cM_H$ has the lowest possible dimension and
thus this case represents a nice laboratory to study quantum corrections, gaugings, fluxes, etc.
(see, for instance,
\cite{Strominger:1997eb,Gutperle:2000sb,Ceresole:2001wi,Antoniadis:2003sw,Davidse:2004gg,Kachru:2004jr,Bao:2009fg,Catino:2013syn}).
Moreover, the metric on four-dimensional QK spaces allows an explicit parametrization
\cite{Przanowski:1991ru,MR1423177}, which reduces it to a solution of an integrable system.
In particular, in the presence of one continuous isometry, it is encoded in a solution of
the integrable {\it Toda} equation. This fact was extensively used in several studies of instantons and their
impact on moduli stabilization
\cite{Ketov:2001ky,Ketov:2001gq,Ketov:2002vr,Davidse:2005ef,Alexandrov:2006hx,Alexandrov:2012np}.

Here we use the explicit results of \cite{Alexandrov:2014sya} providing the {\it exact} metric on $\cM_H$
corrected by D-instantons with mutually local charges,
which was shown to be consistent with the description based on the Toda equation.
As explained in the end of the previous subsection, the $H$-fluxes
protect one linear combination of the isometries $k$ and $\tlk$.
Since in the rigid case the D-instanton charge is a two-dimensional vector,
$\gamma=(p,q)$, the charges of the allowed D-instantons are necessarily mutually local.
Thus, the metric computed in \cite{Alexandrov:2014sya} contains {\it all}\; instantons allowed by the fluxes.

Explicitly, this metric is given by
\be
\de s^2=
\frac{2}{r^2}\[\(1-\frac{2r}{\cR^2\Uin}\) \((\de r)^2+\frac{\cR^2}{4}\,|\cY|^2\)
+\frac{1}{64}\(1-\frac{2r}{\cR^2\Uin}\)^{-1}\(\de \sigma +\tzeta \de \zeta-\zeta\de \tzeta+\cVs \)^2\],
\label{mett-UHMmain}
\ee
where all notations, such as $\cR$, $\Uin$, $\cY$, $\cVs$, are explained in Appendix \ref{subap-metric}.
The charge vectors \eqref{ch-gauge} corresponding to our choice of fluxes
are given by
\be
\begin{split}
\bfk_0=&\,
\thh\p_{\tzeta}+h\p_\zeta +\(2e_0+h\tzeta-\thh\zeta\) \p_\sigma\, ,
\\
\bfk_i=&\,
2e_i\p_\sigma\, .
\end{split}
\label{kilv-H}
\ee
They generate isometries of the metric \eqref{mett-UHMmain}
provided that the D-instanton charges are restricted to satisfy  $hq=\thh p$.
The associated moment maps $\vec\mu_I$ are computed in Appendix \ref{subap-moment} with the following result:
\be
\begin{split}
\mu_i^+=&\, 0,
\qquad\qquad\qquad\quad
\mu_i^3=\frac{e_i}{2r},
\\
\mu_0^+=&\, \frac{\I\cR}{2r}\(\thh-\lambda h\),
\qquad
\mu_0^3=\frac{1}{2r}\(e_0+h\tzeta-\thh\zeta\).
\end{split}
\label{momentmap-main}
\ee
Thus, the only effect of instantons on the moment maps is contained
in the function $\cR$ determined by the equation \eqref{r-UHM}.

Now we use all these data to compute the scalar potential \eqref{scpot-gen}.
A simple calculation gives
\be
\begin{split}
V=&\, \frac{e^{\cK}}{4r^2}\Biggl[ \frac{2|E+\cE|^2}{1-\frac{2r}{\cR^2\Uin}}
+\cK^{i\bj}\(e_i+E\cK_i\)\(e_j+\bE\cK_{\bj}\)-3|E|^2
+4\cR^2|\thh-\lambda h|^2\(\cK^{i\bj}\cK_i\cK_{\bj}-1-\frac{4r}{\cR^2\Uin}\)
\Biggr],
\end{split}
\label{potential-main}
\ee
where $\cK_i=\p_i\cK$ and we have denoted
\be
\begin{split}
E=&\, e_0+h\tzeta-\thh\zeta+e_i z^i,
\\
\cE=&\, \hf\(h\iota_{\p_\zeta}+\thh\iota_{\p_{\tzeta}}\)\cVs.
\end{split}
\label{defE}
\ee

Note that both the metric and the potential are invariant under
the symplectic transformations induced by a change of basis of 3-cycles on $\CY$.
This invariance can be used to put $h$-flux to zero, which we assume from now on.
In this symplectic frame, only electrically charged instantons contribute to the potential.
Using this simplification, one can show that
\be
\cE=\frac{4\thh r\bvl}{\cR (|\Min|^2+|\vl|^2)}\, ,
\ee
where the quantities appearing on the r.h.s., initially introduced in Appendix \ref{subap-metric},
can now be computed explicitly as
\bea
\vl &=& 384 c\sum_{q>0}s(q) q^2 \sin(2\pi q\zeta)K_1(4\pi q\cR),
\nn\\
\Min &=& 2\lambda_2+384 c\sum_{q>0}s(q) q^2 \cos(2\pi q\zeta)K_0(4\pi q\cR),
\label{res-electric}\\
r &=& \frac{\lambda_2\cR^2}{2}-c-\frac{24c\cR}{\pi}\sum_{q>0}s(q) q \cos(2\pi q\zeta)K_1(4\pi q\cR),
\nn
\eea
whereas $\Uin$, also appearing in the potential \eqref{potential-main}, is still given by \eqref{Ab-UHM}.
Here we have introduced the divisor function
\be
s(q)\equiv\sigma_{-2}(q)=\sum_{d|n}d^{-2},
\ee
and, using \eqref{Omq} and \eqref{def-c}, expressed the DT invariants, counting the D-instantons, via the parameter $c$.
As a result, all $g_s$-corrections affecting the scalar potential
are controlled by just one topological number! It is in contrast to the
$\alpha'$-corrections which require knowledge of an infinite set of genus-zero
Gopakumar-Vafa invariants.

\section{Moduli stabilization}
\label{sec-modstab}

Given the scalar potential \eqref{potential-main},
we can investigate whether it has local minima where all moduli are stabilized.
If such minima exist, the sign of the potential evaluated at these points indicates
whether they correspond to a de Sitter or an anti-de Sitter vacuum.

At $h=0$ the potential explicitly depends on dilaton $r$, K\"ahler moduli $t^i$, periods $b^i$ of the
$B$-field, and the RR scalar $\zeta$, and is independent of another RR scalar $\tzeta$ and the
NS-axion $\sigma$. This fact, however, is not a problem for moduli stabilization
since these are the scalars which are used for the gauging.
In the effective action, one can redefine some of the gauge fields to absorb these scalars.
In such frame the scalars are ``eaten up" and thus disappear from the spectrum,
whereas the corresponding gauge fields become massive.

It is also important to note that in the perturbative approximation
the potential depends on the fields $b^i$ and $\zeta$, known
as {\it axions},\footnote{The axions also include the ``eaten up" fields $\tzeta$ and $\sigma$.}
only through the combination $e_i b^i-\thh\zeta$ appearing in \eqref{defE}.
Thus, the other $h^{1,1}$ independent combinations of these fields enter the potential only via instanton corrections:
$b^i$ and $\zeta$ appear in the imaginary part of the worldsheet and the D-instanton actions, respectively.
This shows that the instanton corrections are indispensable for stabilization of all moduli.\footnote{In the given case of rigid CY,
this argument does not allow us to conclude that D-instantons are truly necessary,
since worldsheet instantons together with the combination $e_i b^i-\thh\zeta$ lead to a dependence on all axions.
However,  when $h^{2,1}>0$, it is still true that only one combination of RR-scalars appears in
the perturbative potential \cite{Louis:2002ny}, so that D-instantons must be taken into account to stabilize all moduli.}

The instanton corrected potential \eqref{potential-main} leads to a very complicated system of equations on its extrema.
However, if one assumes that the fluxes satisfy the relation
\be
e_0=(n\thh-\ell^i e_i)/2,
\qquad n,\ell^i\in\IZ,
\label{relflux}
\ee
there exists a very simple solution for the axions,
\be
\zeta=n/2,
\qquad
b^i=\ell^i/2.
\label{vanishsol}
\ee
Indeed, using the expressions for the inverse metric $\cK^{i\bj}$ \eqref{invcK}
and the first derivative of the K\"ahler potential $\cK_i$ \eqref{derK},
the scalar potential can be rewritten as
\be
\begin{split}
\hspace{-0.5cm}
V=&\,
\frac{e^{\cK}}{4r^2}\Biggl[ \frac{2|E+\cE|^2}{1-\frac{2r}{\cR^2\Uin}}-2|E|^2
-e^{-\cK}\hN^{ij}e_i e_j
+\frac{\Bigl[\Re\Bigl(E+e^{-\cK}\hN^{kl}\cK_k e_l\Bigr)\Bigr]^2  \!\! +4\thh^2\cR^2}{e^{-\cK}\hN^{i\bj}\cK_i\cK_{\bj}-1}
-\frac{16\thh^2 r}{\Uin} \Biggr] \!,
\end{split}
\label{potall}
\ee
where $\hN^{ij}$ is the inverse of $N_{ij}=-2\Im F_{ij}$.
Besides, it is straightforward to verify by using the explicit formulae \eqref{res-electric} and \eqref{res-FVM}
that at the point \eqref{vanishsol} all the following quantities vanish:
\be
\Re(E), \quad \vl, \quad \cE, \quad \Re\cK_i, \quad \p_\zeta\cR, \quad
\p_\zeta \Min, \quad \p_\zeta \Uin, \quad \p_{b^i} N_{jk},
\quad \p_{b^i}\cK.
\label{quant-vanish}
\ee
Taking also into account that $\hN^{j\bk}\cK_j\cK_{\bk}=\hN^{jk}\Re\cK_j\Re\cK_{k}+e^{2\cK}N_{ij}t^it^j$,
these results imply that the potential \eqref{potall} satisfies
\be
\left.\p_{\zeta}V\right|_{\zeta=n/2\atop b^i=\ell^i/2}=0,
\qquad
\left.\p_{b^i}V\right|_{\zeta=n/2\atop b^i=\ell^i/2}=0.
\label{bzvac-many}
\ee
Thus, given the fluxes satisfying \eqref{relflux}, half-integer axions
are always a solution of (at least, half of) the equations on critical points.

Of course, there is no guarantee that sticking to this solution would allow to stabilize
the remaining moduli and to get a local minimum, not a saddle point of the potential.
Note, however, that the above properties also imply that the mixed second derivatives vanish,
\be
\left.\p_{\varphi^I}\p_{\psi^J}V\right|_{\zeta=n/2\atop b^i=\ell^i/2}=0,
\label{bzvac-many2}
\ee
where we have introduced the collective notation for the axions, $\psi^I=(\zeta,b^i)$,
and for the remaining fields, $\varphi^I=(r,t^i)$.
This result means that the matrix of the second derivatives has a block-diagonal form,
\be
\p\p V=\( \begin{array}{cc}
\p_{\varphi^I}\p_{\varphi^J}V \ &\ 0
\\
0 \ &\  \p_{\psi^I}\p_{\psi^J}V
\end{array}\),
\label{2derM}
\ee
so that the condition of having a local minimum gives rise to the two independent conditions
on the positive definiteness of $\p_{\varphi^I}\p_{\varphi^J}V$ and $\p_{\psi^I}\p_{\psi^J}V$.
Furthermore, the integers $n$ and $\ell^i$ control the signs of instanton contributions.
One may expect that changing these integers, it may be possible to adjust the signs in such a way that
the matrix $\p_{\psi^I}\p_{\psi^J}V$ becomes positive definite, thus providing a local minimum in
the subspace spanned by the axions, whereas the positive definiteness of $\p_{\varphi^I}\p_{\varphi^J}V$
would impose certain restrictions on the critical points in the remaining subspace.

Thus, in the following, we choose to work with the solution \eqref{vanishsol}.
Having restricted ourselves to this solution, we can significantly simplify the potential.
Using the vanishing of \eqref{quant-vanish}, we find
\be
\begin{split}
\Va(r,t^i)\equiv \left.V\right|_{\zeta=n/2\atop b^i=\ell^i/2}
=&\,
\frac{e^{\cK}}{4r^2}\[ \frac{4r(et)^2}{\cR^2\Min-2r}
-e^{-\cK}\hN^{ij}e_i e_j
+\frac{4\thh^2\cR^2}{e^{\cK}N_{ij}t^i t^j-1}-\frac{16\thh^2 r}{\Min} \].
\end{split}
\label{potallzero}
\ee

Having fixed the axions, we still have to stabilize the four-dimensional dilaton $r$ and the K\"ahler moduli $t^i$.
To this end, we need to solve the equations obtained by variation of the potential \eqref{potallzero}
with respect to these moduli. However, we find it more natural to consider the potential
as a function of $\cR$ rather than of the dilaton $r$ because $\cR(r)$ is defined only implicitly:
see the last equation in \eqref{res-electric} where $\cos(2\pi q\zeta)$
should now be replaced by $(-1)^{nq}$. Proceeding this way and using that
\be
\p_\cR r=\frac{\cR}{4}\(\Min+2\lambda_2\),
\ee
we obtain the following equations:
\begin{subequations}
\bea
\p_\cR \Va
&=&
\frac{e^{\cK}}{4r^2}\Biggl[\frac{\cR}{2r}\(M+2\lambda_2\) e^{-\cK}\hN^{ij}e_i e_j
\Biggr.
\nn\\
&&
-\frac{(et)^2\cR}{\(\cR^2\Min-2r\)^2}\(\cR^2\Min^2+2\lambda_2\(\cR^2\Min-4r\)+4r\(\Min+\cR\p_\cR\Min\)\)
\nn\\
&& \Biggl.
+\frac{2\thh^2\cR\(4r-\cR^2(\Min+2\lambda_2)\)}{r\(e^{\cK}N_{ij}t^i t^j-1\)}
+\frac{4\thh^2}{\Min^2}\(\cR\Min(\Min+2\lambda_2)+4r\p_\cR\Min\) \Biggr]=0,
\label{derR-potallzero}
\\
\p_{t^i} \Va
&=&
-\frac{1}{2r^2}\[ 4r e^{2\cK}\( \frac{(et)}{\cR^2\Min-2r}\( (et)N_{ij}t^j-e^{-\cK}e_i\)-\frac{4\thh^2}{\Min}\,N_{ij}t^j\)
\right.
\nn\\
&& \left.
+\Re F_{ijk}\(\hN^{jm}e_m\hN^{kn} e_n-\frac{4e^{2\cK}\thh^2\cR^2 t^j t^k}{\(e^{\cK}N_{ij}t^i t^j-1\)^2}\) \]=0.
\label{derz-potallzero}
\eea
\label{der-potallzero}
\end{subequations}
Unfortunately, in their full generality, these equations are too complicated for an analytic treatment.
Therefore, they should be studied either numerically or perturbatively.
For instance, we can first analyze them by neglecting all non-perturbative corrections,
and then add the terms with worldsheet and D-brane instantons.
In the next section, we perform the first step, and then in section \ref{sec-inst} attempt
the second step in the special case $h^{1,1}=1$.

It is important to note that the fields to be stabilized cannot take arbitrary values,
being restricted to certain physical domains. These restrictions typically appear due to various
approximations used to get the scalar potential, while approaching a boundary of a physical domain
corresponds to a failure of one of such approximations. The physical domains are defined by the following conditions:
\begin{itemize}
\item
The K\"ahler moduli $t^i$ must belong to the K\"ahler cone of $\CY$ and be
such that the K\"ahler potential is well defined, which implies that $e^{-\cK}>0$.
This quantity is explicitly computed in \eqref{resK}.
Typically, its positivity is ensured by the instanton contributions,
but in the perturbative approximation with $t^i$ sufficiently small,
one can reach a point where the negative perturbative correction becomes dominant
over the classical volume term. This indicates the breakdown of the perturbative approximation
and puts a bound on the domain of the K\"ahler moduli.

\item
Similarly, the K\"ahler moduli must be such that $\Im\cN_{IJ}$, defined in \eqref{defcN}
and determining the kinetic terms of the gauge fields, and its inverse computed in \eqref{relNN},
are negative definite.

\item
The four-dimensional dilaton $r=e^\phi$, besides being positive, should satisfy an additional bound.
In \cite{Alexandrov:2014sya} it was shown that the metric \eqref{mett-UHMmain} has a curvature singularity at the hypersurface
determined by the equation $r=\hf\, \cR^2\Uin$. However, the metric on the physical moduli space must be regular.
Thus, the curvature singularity is an artefact of an approximation: in the case of fluxless CY compactifications,
it is believed that it should be resolved by NS5-brane instantons \cite{Alexandrov:2009qq},
whereas in our case it should probably disappear after taking into account the back reaction of fluxes.
This implies that close to the singularity the metric \eqref{mett-UHMmain} and, hence,
the corresponding scalar potential cannot be trusted. In other words, we should require that $r>r_{\rm cr}$.
In the perturbative approximation one has $r_{\rm cr}=-2c$.
\end{itemize}

\section{Perturbative approximation}
\label{sec-pert}

After dropping all instanton corrections, the scalar potential \eqref{potallzero} takes the following form:
\be
\Va\approx \frac{e^{\cK}}{8r^2}\[ \frac{16\thh^2}{\lambda_2}\,\frac{(1-\gp)r+2c}{1+\gp}
+\frac{4r(et)^2}{r+2c}-e^{-\cK}\kappa^{ij}e_i e_j \],
\label{pertpot}
\ee
where the sign $\approx$ means that the equation holds in the perturbative approximation,
$\kappa^{ij}$ is the inverse of $\kappa_{ij}\equiv\kappa_{ijk}t^k$, and we have introduced
\be
\gp= 3 C e^\cK=\frac{3\chi_\CY}{4\pi^3}\, \zeta(3) e^\cK
\label{deggp}
\ee
as the variable encoding the volume $\cV$ of the Calabi-Yau space since $e^{-\cK}\approx 8\cV-C$ due to \eqref{resK}.
Note that both $\kappa^{ij}$ and $\gp$ are functions of the K\"ahler moduli.

The equations on critical points \eqref{der-potallzero} simplify as
\begin{subequations}
\bea
e^{-\cK}\kappa^{ij}e_i e_j&\approx &
\frac{4(et)^2 r(r+c)}{(r+2c)^2}+ \frac{8\thh^2}{\lambda_2}\,\frac{(1-\gp)r+4c}{1+\gp}\, ,
\label{eqdil-pertall}
\\
\kappa_{ijk}\kappa^{jm}e_m\kappa^{kn}e_n &\approx &
\frac{8r e^\cK(et)}{r+2c}\(2e^\cK (et)\kappa_{ij}t^j -e_i\)
+\frac{64\thh^2}{\lambda_2}\, e^{2\cK}\kappa_{ij}t^j\(\frac{2(r+c)}{(1+\gp)^2}-r\).
\label{stabK-allpert}
\eea
\label{eq-pertall}
\end{subequations}
The main complication here comes from the presence of the inverse matrix $\kappa^{ij}$ that introduces
a non-polynomial dependence on the K\"ahler moduli. It is, however,  possible to get at least one equation without such dependence.
To this end, let us contract \eqref{stabK-allpert} with $t^i$. This gives
\be
e^{-\cK}\kappa^{ij}e_i e_j\approx
\frac{4r (et)^2(1+\gp)}{r+2c}
+\frac{16\thh^2}{\lambda_2}(3+\gp)\(\frac{2(r+c)}{(1+\gp)^2}-r\).
\label{proj-stabKt}
\ee
Combining this equation with \eqref{eqdil-pertall} leads to
\be
\frac{r(et)^2}{(r+2c)^2}\approx
\frac{2\thh^2}{\lambda_2}\,\frac{\(2\gp^3+9\gp^2 +10\gp-5\)r-8c}{(1+\gp)^2\(\gp(r+2c)+c\)}\, ,
\label{eq-et}
\ee
which is a cubic equation on the dilaton $r$.
Furthermore, substituting \eqref{eqdil-pertall} and \eqref{eq-et} into the perturbative potential \eqref{pertpot},
we find the following result for its value at critical points:
\be
\begin{split}
\Va_{\rm cr}
\approx &\,
\frac{e^\cK}{r}\[ \frac{\thh^2}{\lambda_2}\,\frac{1-\gp}{ 1+\gp}
+\frac{c(et)^2}{2(r+2c)^2}\]
\\
\approx &\,
\frac{e^\cK\thh^2}{\lambda_2 r^2}\, \frac{\gp(1-\gp^2)r^2-4c(1-3\gp-2\gp^2) r-8c^2 }{(1+\gp)^2\(\gp(r+2c)+c\)}\, .
\end{split}
\label{extvalue}
\ee

In principle, one can solve the cubic equation \eqref{eq-et} to express $r$ in terms of the combination $e_it^i$
and the Calabi-Yau volume encoded in $\gp$. The solution $r(t)$ is to be substituted into \eqref{stabK-allpert},
which leads to a complicated system of equations on the K\"ahler moduli.
But even without explicitly solving this system,
it turns out to be possible to derive some bounds on its solution.

\subsection{Bounds on perturbative solutions}
\label{subsec-bounds}

As we noticed in the end of section \ref{sec-modstab}, the possible values of the scalar fields are restricted
to satisfy certain conditions. In the perturbative approximation, two of them put simple bounds on the lowest values
of the dilaton (inversely proportional to the string coupling) and the volume of CY,
\be
r>2|c|,
\qquad
\cV>C/8,
\label{bound-rV}
\ee
whereas the third one demands that $\Im\cN_{IJ}$ is negative definite.
The last condition is equivalent to $\Im\cN^{IJ}v_Iv_J<0$ for any real vector $v_I$.
Let us take $v_I=(-(eb),e_i)$. Then, using the perturbative result \eqref{pert-cNinv},
we arrive at the following condition:
\be
\frac{e^{-\cK}\kappa^{ij}e_i e_j}{(et)^2}<4.
\label{condN}
\ee

Let us now apply this condition to the extrema of the potential.
Using equations \eqref{eqdil-pertall} and \eqref{eq-et}, we find that
\be
e^{-\cK}\kappa^{ij}e_i e_j-4(et)^2\approx
\frac{8\thh^2}{\lambda_2 r}\,\frac{\gp(1-\gp^2)r^3+8c\(2-3\gp-3\gp^2-\gp^3\)r^2+4c^2\(12-7\gp-7\gp^2-2\gp^3\) r+32 c^3}
{(1+\gp)^2\(\gp(r+2c)+c\)}.
\label{exact-cond}
\ee
Then \eqref{condN} implies that the r.h.s. of \eqref{exact-cond} must be negative. This severely restricts
the regions in the $\gp$-$r$ plane where the potential can have critical points. Furthermore, the positivity of
\eqref{eq-et} gives another condition of the same kind. Fig. \ref{fig-region} shows the regions allowed by the two conditions,
as well as those where the potential \eqref{extvalue} is positive.
We observe that there is a narrow region where all conditions are satisfied so that they do
not exclude the existence of meta-stable dS vacua, although they put a strong upper bound on the dilaton.

\twofig{The plane $\gp$-$(r/|c|)$ and its regions where various conditions are satisfied:
the condition (4.8) corresponds to the dark grey region with the blue boundary, positivity of (4.5)
holds in the pink region with the purple boundary, and the potential at the extremum (4.6)
is positive in the light grey region with the brown boundary.
The right picture magnifies the region close to the bifurcation point corresponding to
$
\(\gp_\star=\frac14\, (\sqrt{17}-3),r_\star=\frac{|c|}{2}\, (\sqrt{17}+7 )\).
$
All three conditions are satisfied only in the very narrow region which ends at this point.
If one drops the positivity of the potential, the region of large $\gp$ and $r$ is also allowed.}
{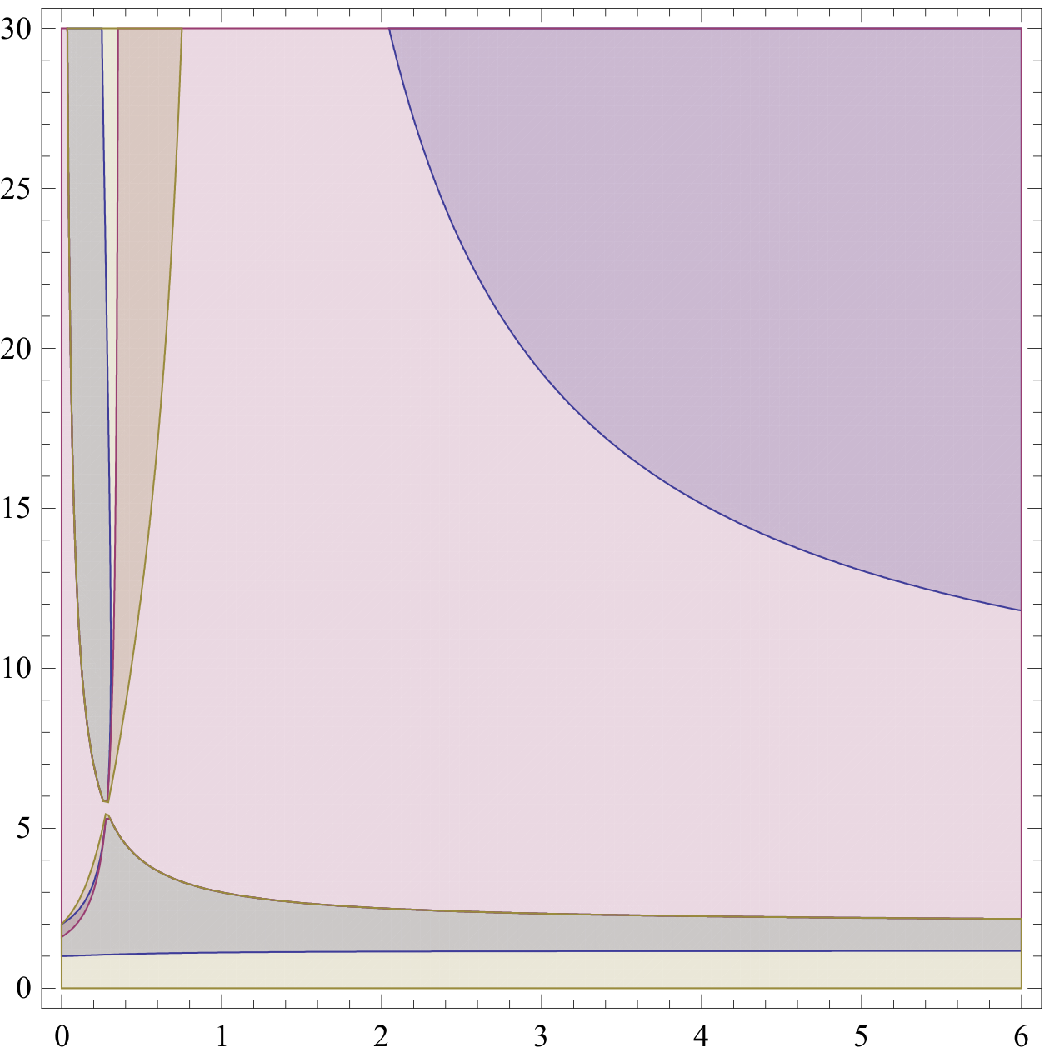}{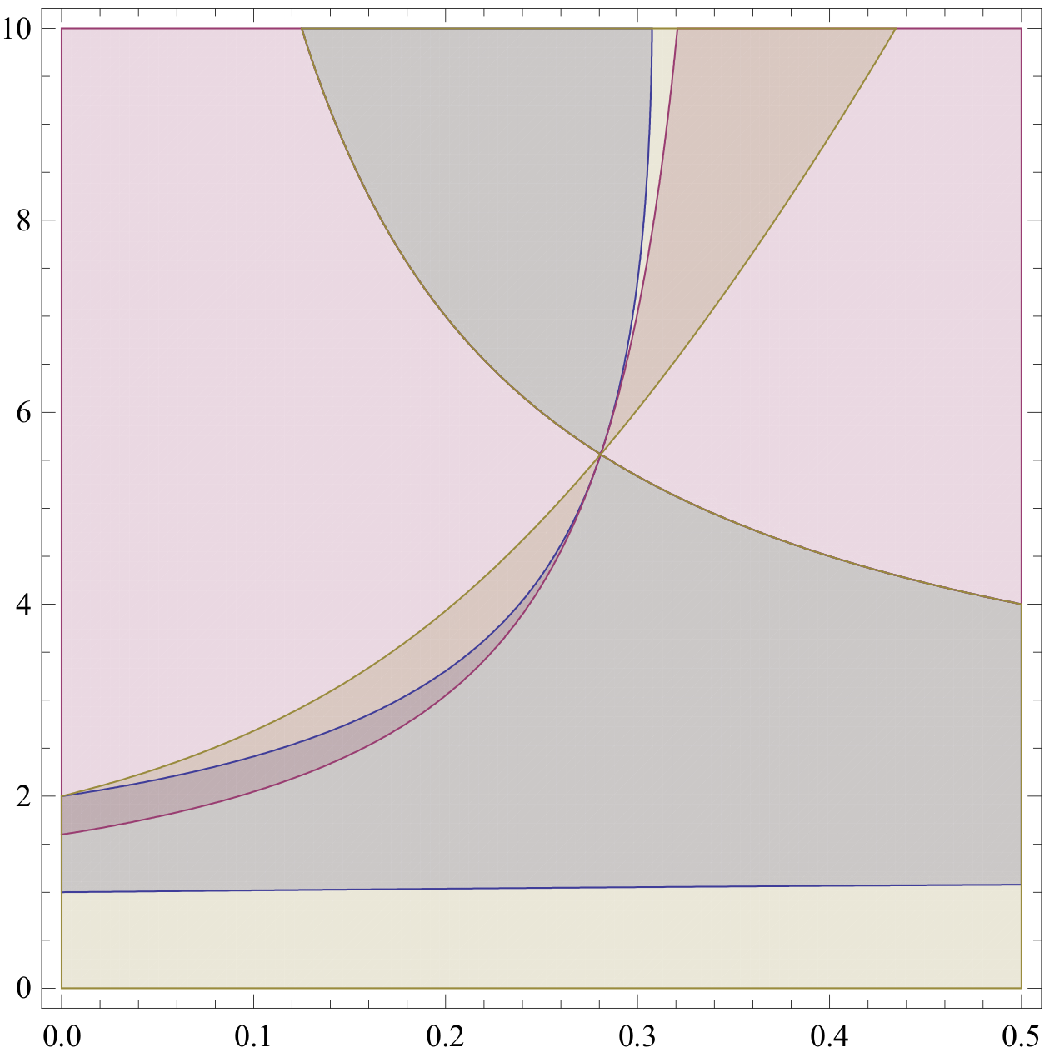}{8.5cm}{fig-region}{0.1cm}{0.1cm}

It is the important feature of our results presented in Fig. \ref{fig-region} that
the above conditions do {\it not} allow solutions which have {\it both} large $r$ (small string coupling)
and small $\gp$ (large volume). Such conclusion can actually be derived analytically.
Indeed, it is enough to get a milder consequence of \eqref{condN} than the negativity of \eqref{exact-cond}.
For instance, one can note that the first term in \eqref{eqdil-pertall} is larger than $4(et)^2$. Then \eqref{condN}
implies that the second term must be negative, which is equivalent to
\be
\gp > 1+\frac{4c}{r}
\quad \Rightarrow \quad  \(1+\frac{c}{r}\)\(1-\frac{C}{8\cV}\)<\frac{3}{4}\, .
\label{second-cond2}
\ee
When both $r$ and $\cV$ are large, which corresponds to the classical limit, this condition is clearly violated.
This shows that for the set of fluxes under consideration, it is impossible to stabilize the string coupling and the volume
in the region where all quantum corrections become irrelevant. Note also that the bound does not depend on the values of fluxes,
which means that it is impossible to tune them in order to get arbitrarily large $r$ and $\cV$.

\subsection{One-modulus case}
\label{subsec-one}

Given a complicated structure of the equations on critical points of the scalar potential even in the perturbative approximation,
it is natural to consider some particular cases with a low number of moduli. First, we concentrate on the simplest case
with a single K\"ahler modulus, corresponding to CY with Hodge numbers $(h^{1,1},h^{2,1})=(1,0)$. To the best of
our knowledge, no CY manifolds with such topological characteristics have been constructed so far,
so that this case represents a fictional geometry and the corresponding gauged supergravity has no direct connection
to string theory.
Nevertheless, it is instructive to study it because the resulting equations allow an analytic treatment.

In the one modulus case, we find an additional relation,
\be
\frac{e^{-\cK}\kappa^{ij} e_i e_j}{(et)^2}\approx \frac{4}{3+\gp}\, ,
\label{onemodN}
\ee
where $\kappa^{ij} e_i e_j=\frac{e_1^2}{\kappa_{111}t^1}$. It allows to rewrite the cubic equation on the dilaton
in the form where the coefficients are functions of $\gp$ only. Namely, combining \eqref{eqdil-pertall}, \eqref{eq-et}
and \eqref{onemodN}, we find
\be
(5+\gp)(2-4\gp-5\gp^2-\gp^3)r^3+4(2+\gp-4\gp^2-\gp^3)cr^2-8(5-\gp)c^2r-32c^3=0.
\ee
Remarkably, this equation can be factorized so that all three roots can be found explicitly as
\be
\begin{split}
r_0 =&\, \frac{4|c|}{5+\gp}\, ,
\qquad
r_\pm = \frac{2|c|\(\gp\pm \sqrt{(2+\gp)(2-5\gp-2\gp^2)}\)}{2-\gp(1+\gp)(4+\gp)}\, .
\end{split}
\label{roots}
\ee
However, not all of them are relevant to us. First, we observe that $r_0<|c|$ and,
hence, this root violates the bound \eqref{bound-rV}. The other two roots are real only when
\be
\gp<\gp_{(1)}=\frac14\(\sqrt{41}-5\)\approx 0.3508.
\label{r-bound1}
\ee
However, in this region we have $r_-<0$. Thus, only $r_+$ should be considered,
whose positivity puts a stronger bound than \eqref{r-bound1},
namely,\footnote{$\gp_{(2)}$ is one of the roots of the denominator in \eqref{roots}.}
\be
\gp<\gp_{(2)}\approx 0.3429.
\label{r-bound2}
\ee
We should also check the two conditions mentioned in the previous subsection: \eqref{condN}
and positivity of \eqref{eq-et}. The first one is automatically satisfied due to \eqref{onemodN}, whereas
the second one leads to an even stronger bound,\footnote{$\gp_\star$ is one of the roots of the denominator in \eqref{eq-et} after
substitution $r=r_+$, namely, it solves $\gp(r_+(\gp)+2c)+c=0$.
It coincides with the bifurcation point in Fig. \ref{fig-region}, which is independent of the number of moduli.
Note also that for $\gp_\star<\gp<\gp_{(2)}$,
the r.h.s. of \eqref{exact-cond} is positive, which seems to contradict to \eqref{condN}.
In fact, there is no contradiction because in this domain one already violates the bound \eqref{r-bound3}.}
\be
\gp<\gp_\star\approx 0.2808,
\label{r-bound3}
\ee
where $\gamma_\star$ was defined in the caption to Fig. \ref{fig-region}.

\twofig{The graphs represent the same quantity $\frac{\lambda_2}{|c|\thh^2}(et)^2|_{r=r_+(\gp)}$
evaluated in the one-modulus case as a function of the parameter $\gp$ in the two ways:
the blue curve represents the function (4.5) and the red curve represents the function
$f^{-1}(1+3\gp^{-1})^{2/3}\sim t^2$ obtained by using (4.17).
The parameter $f$ controls the height of the second curve.
For large $f$ the curves intersect at two points (the left picture with $f=26$)
corresponding to two extrema of the potential, whereas for small $f$ there are no intersections (the right picture with $f=6.5$).}
{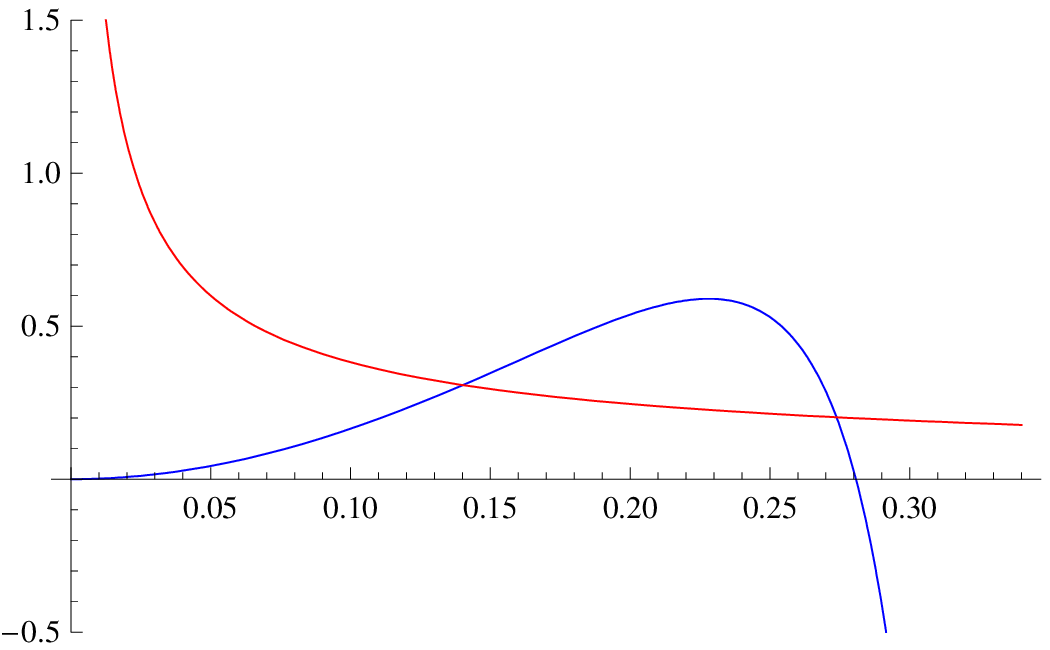}{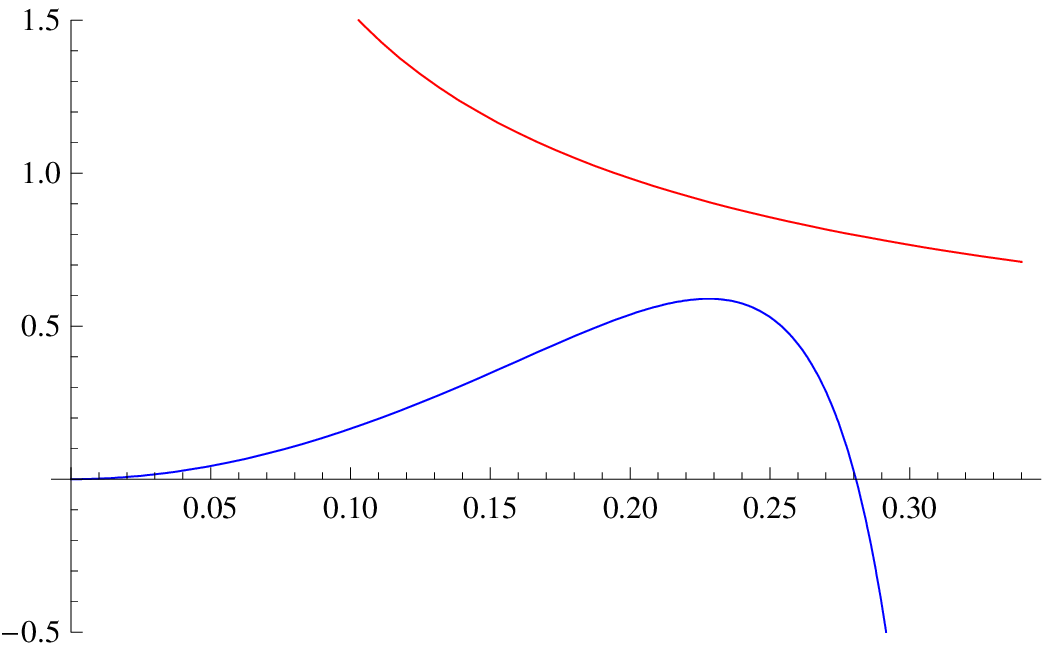}{9cm}{fig-solve}{0.0cm}{-0.0cm}

Having verified all our conditions, it remains to solve the equation fixing the modulus $\gp$.
The easiest way to obtain such equation is to take \eqref{eq-et}, where one should substitute $r=r_+(\gp)$ and
\be
t^1=\(\frac{3C}{4\kappa_{111}}\(1+3\gp^{-1}\)\)^{1/3}.
\label{t-one}
\ee
Unfortunately, a solution can be found only numerically, and it is controlled by the parameter
\be
f=\frac{|c|\thh^2}{\lambda_2 e_1^2}\(\frac{4\kappa_{111}}{3C}\)^{2/3}
=\frac{\pi\thh^2}{24\lambda_2 e_1^2}\(\frac{\kappa_{111}}{3\zeta(3)}\)^{2/3},
\label{param-f}
\ee
where we have used that we are considering the case with $\chi_\CY=2$. One can show that for
\be
f >f_{\rm crit} \approx 9.8
\label{f-bound}
\ee
the equation always has two solutions, and does not have any in the opposite case.
The situation is demonstrated in Fig. \ref{fig-solve} which represents the two sides of Eq. \eqref{eq-et} as
functions of $\gp$. For the parameters satisfying \eqref{f-bound}, the two curves have two intersection points,
but once $f$ decreases and reaches the critical value, they do not intersect anymore.

\lfig{The profile of the potential on the plane $\gp$-$(r/|c|)$.
There is a local maximum at $\gp\approx 0.27$, $r\approx 5.18|c|$ and a saddle point at
$\gp\approx 0.14$, $r\approx 2.66|c|$.
The profile corresponds to the choice $f=26$,
and the potential is rescaled by the factor $\frac{3\lambda_2 |c|C}{\thh^2}$.}
{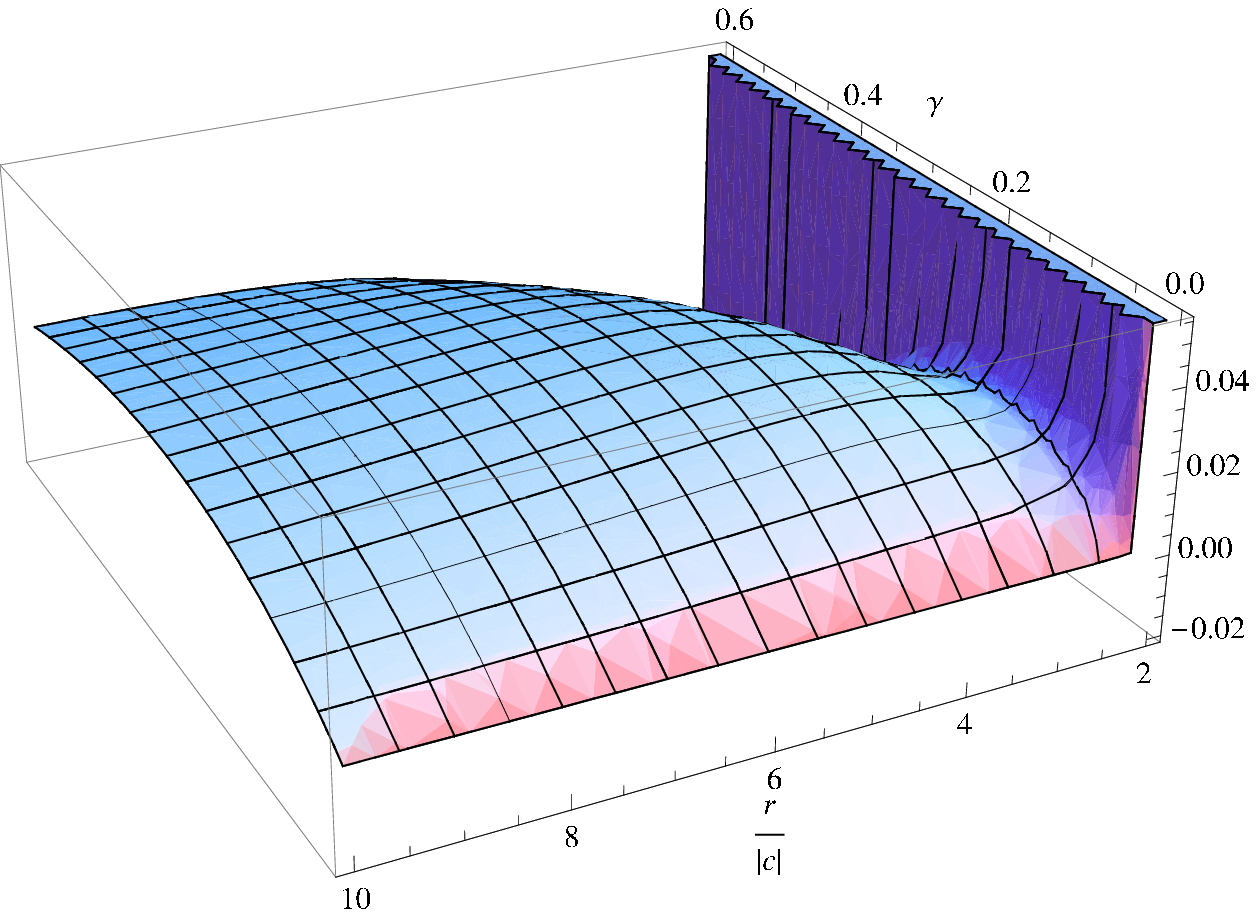}{9.5cm}{fig-potential}{-0.2cm}

Thus, if the $H$-flux is sufficiently large compared to the $F_4$-flux, the potential has two critical points.
Remarkably, for both of them the potential turns out to be positive (the curve $r_+(\gp)$
drawn on the $\gp$-$r$ plane precisely fits the narrow region identified in Fig. \ref{fig-region}).
Unfortunately, both critical points do not correspond to local minima. As can be seen in Fig. \ref{fig-potential},
the solution with larger $\gp$ and $r$ corresponds to a local maximum,
whereas the one with smaller parameters corresponds to a saddle point.
This is also confirmed by our analysis of the matrix of the second derivatives
of the potential performed in Appendix \ref{ap-second}.
As a result, we conclude that in the one-modulus case the perturbative potential does not have
meta-stable vacua.

\subsection{Generic case: stability analysis}
\label{subsec-two}

Next, it is natural to analyze the case with two K\"ahler moduli. Remarkably, a Calabi-Yau manifold with Hodge numbers
$(h^{1,1},h^{2,1})=(2,0)$ was constructed a few years ago in \cite{Freitag:2011}.
Thus, in contrast to the one-modulus case, this one does have a mathematical realization.
The intersection numbers of this CY were recently calculated
in \cite{Freitag:2015}, and are given by\footnote{We are very grateful to Eberhard Freitag for informing us about his calculations.}
\be
\kappa_{111}=344,
\qquad
\kappa_{112}=492
\qquad
\kappa_{122}=600,
\qquad
\kappa_{222}=440.
\label{internum}
\ee
Unfortunately, these numbers do not have any particular symmetry which could help us in solving our equations.
Furthermore, although it is possible to explicitly invert the $2\times 2$ matrix $\kappa_{ij}$ entering these equations,
they still remain unsuitable to an analytic treatment.

Due to these reasons, instead of solving the equations on critical points, we directly proceed to the analysis of meta-stability.
Remarkably, it turns out that this analysis can be carried out for the general case with {\it any} number of K\"ahler moduli.

The meta-stability of a vacuum requires that the matrix of the second derivatives $\p_{\varphi^I}\p_{\varphi^J}\Va$
at the corresponding critical point is positive definite.
To understand whether this can be the case for our potential, we apply the following trick.
First, we note that the signature of any linear operator does not depend on the choice of a basis in the space where it acts.
Therefore, we can rotate the derivatives $\p_{t^i}$ by an invertible matrix ${\mb_i}^j$.
We choose
\be
{\mb_1}^j=t^j,
\qquad
{\mb_2}^j=n^j\equiv \frac{\kappa^{jk}e_k}{e^{\cK}(et)}\, ,
\ee
and ${\mb_i}^j$ with $i>2$ such that together with $t^j$ and $n^j$ they form a set of linearly independent vectors.
Thus, instead of $\p_{\varphi^I}\p_{\varphi^J}\Va$, we are going to analyze a matrix of the following form:
\be
\Mb=\(\begin{array}{cccc}
\p_r^2 \Va & t^i\p_{t^i}\p_r \Va & n^i\p_{t^i}\p_r \Va \ &\
\\
t^i\p_{t^i}\p_r \Va\ &\ t^i t^j\p_{t^i}\p_{t^j} \Va\ &\ n^i t^j\p_{t^i}\p_{t^j} \Va\ &\ \cdots\
\\
n^i\p_{t^i}\p_r \Va\ &\ n^i t^j\p_{t^i}\p_{t^j} \Va\ &\ n^i n^j\p_{t^i}\p_{t^j} \Va\ &\
\\ & \cdots & \ &\ \cdots\
\end{array}\).
\label{matder}
\ee

Since $\Mb$ is a Hermitian matrix, we can apply Sylvester's criterion which tells us that $\Mb$ is positive definite {\it if and only if}
all its leading principal minors are positive. In other words, all matrices $\Mbi{k}$ given by the upper left $k$-by-$k$ corner of $\Mb$
must have a positive determinant, i.e. $\Delta_k\equiv\det\Mbi{k}>0$.
In particular, a {\it necessary} condition for $\Mb$ to be positive definite is the positivity of $\Delta_k$,
$k=1,2,3$.\footnote{In the following, whenever $\Delta_k$ is mentioned, the condition $k=1,2,3$ is implied.}

The crucial fact is that it is possible to express all elements of the matrix $\Mbi{3}$, and hence $\Delta_k$,
in terms of $\gp$ and $r$ only. Indeed, contracting the vector-like equation \eqref{stabK-allpert} with $n^i$, we obtain
\bea
&&
\kappa_{ijk}\,\kappa^{il}e_l\,\kappa^{jm}e_m\,\kappa^{kn}e_n \approx
\frac{8r e^{2\cK}(et)}{r+2c}\(2(et)^2- e^{-\cK} \kappa^{ij}e_ie_j\)
+\frac{64\thh^2}{\lambda_2}\, e^{2\cK}(et)\(\frac{2(r+c)}{(1+\gp)^2}-r\)
\label{proj-stabKn}\\
&&\approx
\frac{32\thh^2}{\lambda_2}\, \frac{(et)e^{2\cK}}{r+2c}\,\frac{ (5-10\gp-13\gp^2-2\gp^3)r^3-2c(1-8\gp+3\gp^2)r^2
-4c^2 (9-8\gp)r-8c^3(3-2\gp)}{(1+\gp)^2\(c+\gp(r+2c)\)},
\nn
\eea
where we have used \eqref{proj-stabKt} and \eqref{eq-et} to get the second line.
Then, as shown in Appendix \ref{ap-second}, using the equations \eqref{proj-stabKt}, \eqref{eq-et} and \eqref{proj-stabKn},
we can express all independent structures appearing in the entries of $\Mbi{3}$ in terms of only two variables $\gp$ and $r$.
As a result, it becomes possible to search for regions in the $\gp$-$r$ plane
where $\Delta_k$ are all positive.
It is important to emphasize that, due to the use of the equations on critical points, all the parameters $\lambda_2$, $\kappa_{ijk}$,
the fluxes $e_i$ and $\thh$ conspire into the same positive multiplicative factor in all entries of the matrix $\Mbi{3}$,
and, hence, in all minors $\Delta_k$, so that the stability analysis does
{\it not} depend on particular values of these parameters.

The details of this analysis are presented in Appendix \ref{ap-second}.
We find that there are no regions in the $\gp$-$r$ plane where all three minors $\Delta_k$ are positive.
This implies that the matrix $\Mb$ \eqref{matder} {\it cannot} be positive definite and, hence,
the perturbative potential {\it cannot} have local minima for any number of K\"ahler moduli.

\section{Instanton contributions in the one-modulus case}
\label{sec-inst}

Given the results of the previous section about the absence of meta-stable vacua in the perturbative approximation,
it is natural to ask whether such vacua exist after taking into account
the non-perturbative corrections generated by worldsheet and D-brane instantons.
In this section we study this question in the simplest case of a fictional CY with $h^{1,1}=1$.
Thus, given all our approximations, the potential analyzed here should be viewed only 
as inspired by string theory, rather than realizing one of its compactifications.
Nevertheless, we expect that it captures the main features of the cases which do have such realization.

In the presence of the non-perturbative corrections it seems to be impossible to solve our equations analytically,
and we have to rely on numerical calculations. Our basic idea is to evaluate
the matrix of the second derivatives of the non-perturbative scalar potential on-shell,
so that the dependence on the flux parameters is completely factorized, and then
look for regions in the $t$-$\cR$ plane\footnote{In this section we drop the index $i$ at
the quantities like K\"ahler moduli since it takes only one value.}
such that
(i) they contain the curve of possible critical points, and (ii) the resulting matrix is positive definite.
More precisely, we perform the following steps:
\begin{enumerate}
\item
Solve \eqref{derz-potallzero}, which in this case is a single equation, with respect to $(et)^2$.
The solution can be represented as
\be
(et)^2=\thh^2 \mE(t,\cR)
\label{et2}
\ee
with some function $\mE(t,\cR)$. Note that this function, as well as all other functions below,
also depend on the signs $(-1)^n$ and $(-1)^l$ determined by the values of the axion fields.
Thus, each function appears in four different copies corresponding to four different choices of these signs.
It is enough to get a local minimum with one of these copies.

\item
Substitute \eqref{et2} into \eqref{derR-potallzero} so that the dependence on $\thh^2$ is factored out
and the equation reduces to
\be
\mQ(t,\cR)=0,
\label{eq-factor}
\ee
where the function $\mQ(t,\cR)$ is independent on the flux parameters.

\item
Calculate the matrix \eqref{2derM}\footnote{More precisely, in the upper left entry
we evaluate the derivatives with respect to $(\cR,\log t)$ instead of $(r,t)$.}
and substitute \eqref{et2}, so that the dependence on $\thh^2$ is also factored out
and the matrix takes the form
\be
\p\p V=\thh^2\( \begin{array}{cc}
\Phi_{IJ}(t,\cR) \ &\ 0
\\
0 \ &\  \Psi_{IJ}(t,\cR)
\end{array}\).
\label{2derV}
\ee

\item
All the steps above can be done analytically. To proceed further, we have to stick to a numerical analysis.
To this end, we fix a finite number of instantons $\Ninst$ to be taken into account, and choose some values for
$\lambda_2$, $\kappa$ and Gopakumar-Vafa invariants $\nk$, $k\le\Ninst$.  We recall that we take a fictional CY,
so that all these numbers can be chosen at will.

\item
Find the low bounds $\cR_{\rm cr}$ and $t_{\rm cr}$ by demanding
\be
r(\cR)> -2c,
\qquad
e^{-\cK(t)}> 0,
\qquad
\Im\cN_{IJ}(t)\ \mbox{is negative definite}.
\label{condnum}
\ee
Under the second condition, the last one can be shown to be equivalent to (see Appendix \ref{ap-cN})
\be
N t^2>e^{-\cK}\qquad \mbox{or}\qquad  N<0,
\label{condNt}
\ee
where $N\equiv N_{11}(t)$.
The subsequent analysis is concentrated on the region $(\cR>\cR_{\rm cr},t>t_{\rm cr})$.

\item \label{mainstep}
Draw the curve $\mQ(t,\cR)=0$ on the $t$-$\cR$ plane, and identify the parts of this curve belonging to the regions where
\begin{itemize}
\item
$\mE(t,\cR)>0$,

\item
the matrix $\Psi_{IJ}(t,\cR)$ is positive definite,

\item
the matrix $\Phi_{IJ}(t,\cR)$ is positive definite.
\end{itemize}

\item
Should such parts exist, it means that there is a range of the flux parameters
that allows the existence of a local minimum of the scalar potential.
This range corresponds to those values of $e$ and $\thh$ when
the two equations, \eqref{et2} and \eqref{eq-factor}, have a common solution.
The fact that such values exist is ensured by positivity of $\mE(t,\cR)$.

\end{enumerate}

\threefigmod{The left picture displays the $t$-$\cR$ plane and its regions where $\mE(t,\cR)>0$ (blue)
and $\Psi_{IJ}(t,\cR)$ is positive definite (pink).
The red curve is a curve of solutions of $\mQ(t,\cR)=0$, while the horizontal and vertical green lines correspond
to $\cR=\cR_{\rm cr}$ and $t=t_{\rm cr}$, respectively. One can see that a part of the red curve belongs to the region
where both conditions are satisfied. The right pictures display the same plane and the curve of
solutions together with the regions of the positive trace (blue) and the determinant (pink) of $\Phi_{IJ}(t,\cR)$.
The lower picture magnifies the part where the two regions are close to each other,
in order to make clear that they do not intersect indeed.
Thus, $\Phi_{IJ}$ is not positive definite near the red curve.
The parameters are chosen as $N_{\rm inst}=4$, $\lambda_2=0.1$, $\kappa=10$, $\nk=100 k$.}
{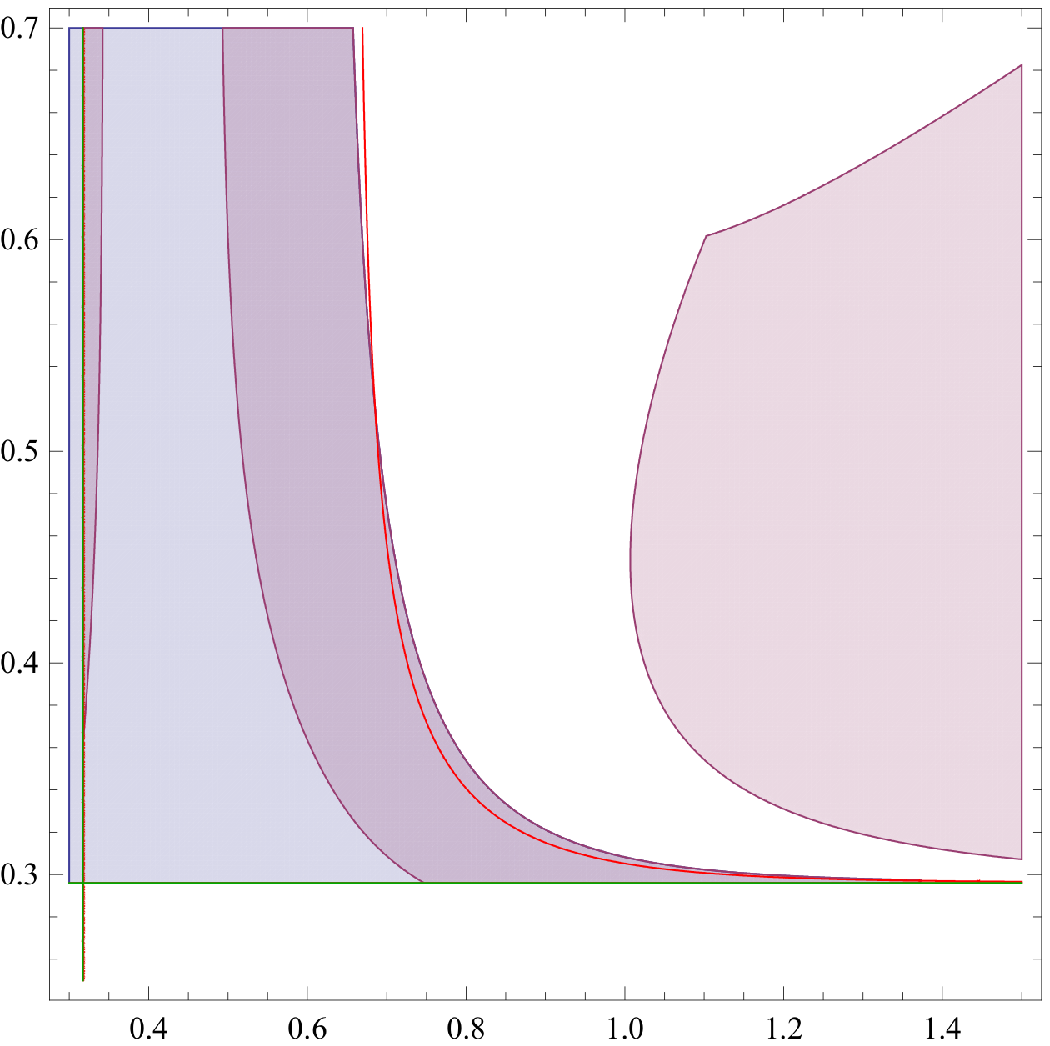}{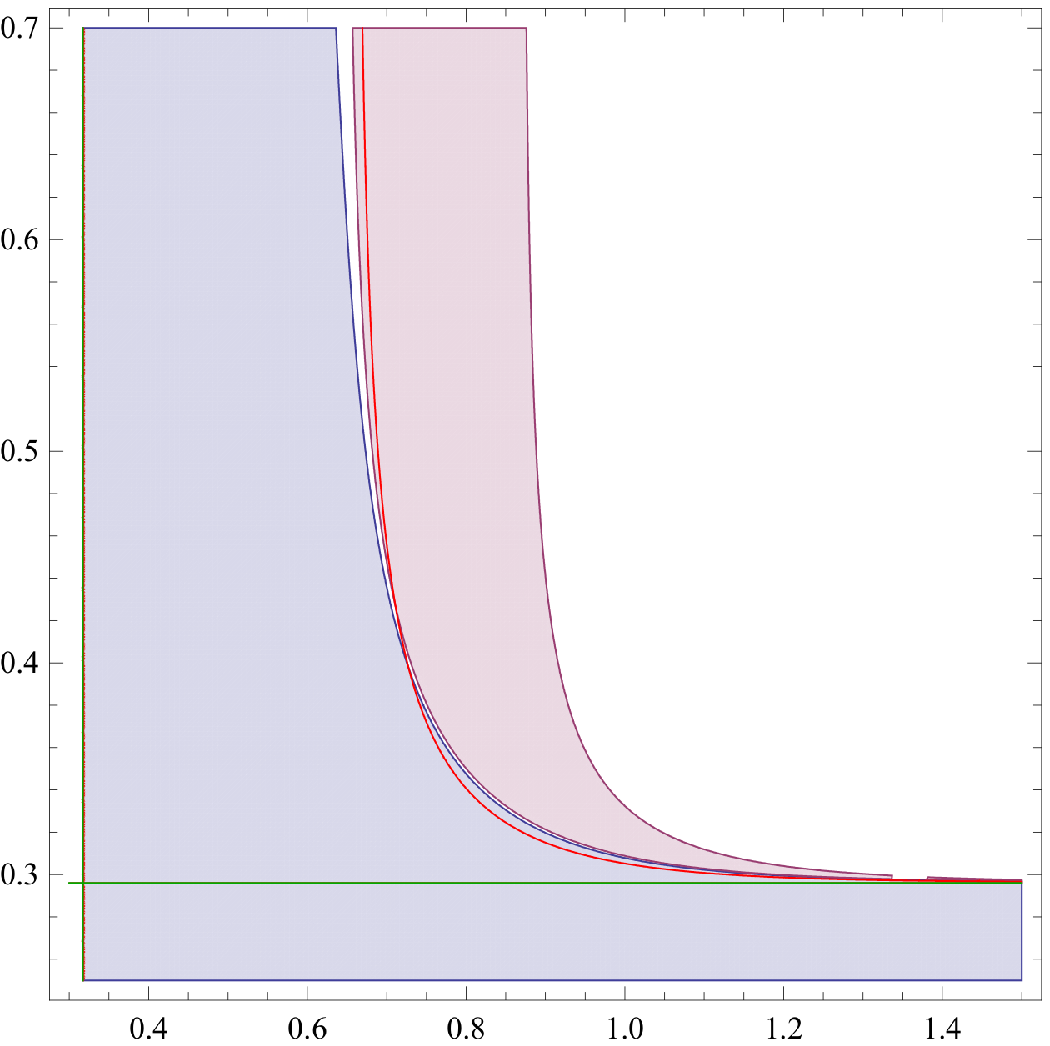}{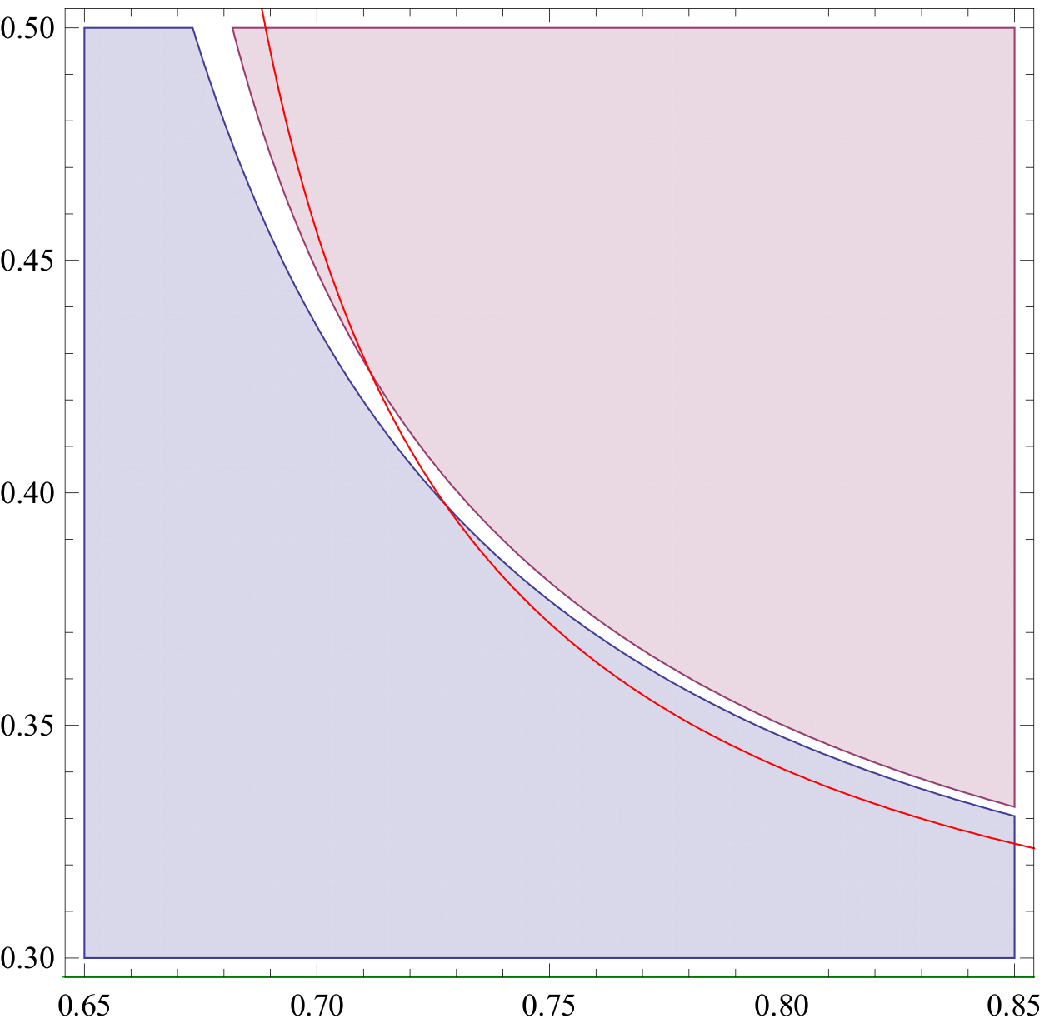}{9cm}{7cm}{fig-inst-regions}{0.2cm}

For practical purposes, it is convenient to split the step \ref{mainstep} into two steps: first, impose
the positivity of $\mE$ and the positive definiteness of $\Psi_{IJ}$, and only afterwards analyze $\Phi_{IJ}$.
Then, typically, at the first stage we can exclude $(-1)^l=1$ and identify a finite part of the curve $\mQ(t,\cR)=0$,
not too far from the critical values, as a candidate for the position of the minima.
However, for all choices of the parameters we considered, it turns out that the matrix $\Phi_{IJ}$
is {\it not} positive definite in the region around the candidate part.
A typical situation is demonstrated in Fig. \ref{fig-inst-regions}.
It is striking that in all our examples the regions of the positive trace
and the positive determinant of $\Phi_{IJ}$ approach each other,
with their boundaries going almost parallel, but {\it never} intersect.
Given a highly non-trivial dependence of these functions on $t$, $\cR$ and all the parameters,
this observation begs for a deeper analytical explanation.\footnote{Actually,
we found that in the deep quantum region (with small $\cR$ and $t$)
it is possible to have $\Phi_{IJ}$ positive definite and the non-perturbative
scalar potential does have local minima.
However, these minima spoil at least one of the conditions \eqref{condnum} and, therefore, are non-physical.}
We conclude that in the one-modulus case the instanton corrections do not lead to meta-stable vacua.

\section{Conclusions}
\label{sec-concl}

In this paper we considered a simple class of flux compactifications which preserve $N=2$ local supersymmetry
in the four-dimensional low energy effective action.
Ignoring the back reaction of fluxes and using the recent results on the non-perturbative description
of fluxless CY compactifications, we derived a scalar potential which takes into account
not only perturbative corrections, but also worldsheet and D-brane instantons.
Extremizing this potential, we found that the axion fields are fixed to half-integer values,
provided the fluxes satisfy the simple constraint \eqref{relflux}. The axion stabilization
greatly simplifies the scalar potential and the equations on its critical points,
and also leads to a factorization of the matrix of its second derivatives,
which allows to disentangle the issue of stability into two independent problems in the subspaces
spanned by the axions and the remaining moduli, respectively.

Whereas the stability in the axion subspace is easy to achieve,
our results on the stabilization of the dilaton and the K\"ahler moduli are largely negative.
First, we found the bound \eqref{second-cond2} on the critical values of
the CY volume and the dilaton, which shows that the scalar potential does not
have critical points in the large volume, weak coupling region of the moduli space
where both $\alpha'$ and $g_s$-corrections can be neglected. Second, we investigated
these critical points in the perturbative approximation, but found that all
of them are {\it not} stable (i.e. not local minima).
Furthermore, in the case with one K\"ahler modulus, corresponding to the non-physical case of rigid CY with $h^{1,1}=1$, 
we extended this result to the non-perturbative level by taking into account all instanton contributions.
Thus, in all these cases not all of the moduli are stabilized by the chosen set of fluxes.
The direction of instability lies in the subspace spanned by the dilaton and the K\"ahler moduli.
This shows the existence of a non-trivial mixture between different moduli,
and a failure of the approximation where
they are supposed to be stabilized in a step-by-step procedure.

Our results can be compared to the no-go theorems in the literature that forbid the existence of dS vacua.
For instance, \cite{Cremmer:1984hj} proves such a theorem in the approximation where the coupling of hypermultiplets is ignored,
i.e. when {\it only} abelian $N=2$ vector multiplets are taken into account, whereas \cite{GomezReino:2008bi}
has a similar statement in the opposite case where {\it only} hypermultiplets are present.
The main differences to these papers are: (i) we take into account {\it both} types of $N=2$ matter multiplets,
and (ii) obtain the stronger result  that not only dS, but {\it any} vacua are unstable.
At the same time, our results only apply either to the perturbative level or to CY's with Hodge numbers $(h^{1,1},h^{2,1})=(1,0)$.

It was argued in \cite{Catino:2013syn} that
meta-stable dS vacua can be obtained in N=2 gauged supergravity with a single hypermultiplet and
a single vector multiplet, by gauging an abelian isometry of the hypermultiplet moduli space. It was based on
the observation that the bound of \cite{GomezReino:2008bi} on (scalar) sGoldstini masses is relaxed in such case.
Our results in the one-modulus case are not in tension with these findings because \cite{Catino:2013syn}
studied the most general  metrics on $\cM_V$ and $\cM_H$, which are consistent with the special
K\"ahler and quaternion-K\"ahler properties, respectively, whereas we restricted
them to those resulting from the fluxless CY compactifications. Rather, our results imply that
the vacua of \cite{Catino:2013syn} are not expected to arise in string theory, at least, 
if the back reaction does not change the situation drastically.

It is also worth mentioning that dS vacua are known to arise after the gauging of {\it non-abelian}
isometries \cite{Fre:2002pd,Ceresole:2014vpa,Fre:2014pca}.
The non-abelian isometries do exist in the {\it classical} supergravity
where the hypermultiplet moduli space can be taken to be a quaternionic homogeneous
space\footnote{For instance, the universal hypermultiplet moduli space,
appearing in CY compactifications in the classical approximation, is given by the symmetric coset space $SU(2,1)/SU(2)\times U(1)$.}
$G/H$ with a semi-simple stability group
$H$, which allows to introduce the so-called Roo-Wagemans angles playing a crucial role
in the construction of the classical dS vacua. However, any {\it quantum} correction,
either perturbative or non-perturbative, breaks the non-abelian symmetries of the hypermultiplet moduli space,
so that the non-abelian gaugings do not apply in quantum
theory.\footnote{That is why we omitted the non-abelian contributions in the basic equation \eqref{scpot-gen}
of the scalar potential in $N=2$ gauged supergravity.} Thus, the vacua constructed in
\cite{Fre:2002pd,Ceresole:2014vpa,Fre:2014pca} do not appear to be relevant in the context of full string theory
where quantum corrections are not ignored.

Returning to our results, we note that they do not fully exclude the class of
flux compactifications which inspired our potential: it remains to understand what happens
at the full non-perturbative level for CY's with $h^{1,1}>1$
(i.e. in all non-fictional cases), and whether the picture we found still persists.
In fact, there is a serious obstacle on this way due to the absence of any knowledge
about Gopakumar-Vafa invariants for rigid CY manifolds. Usually, these invariants are calculated by
using mirror symmetry \cite{Candelas:1990rm,Hosono:1993qy}. However,
rigid CY's do not have mirror duals ($h^{1,1}$ cannot be zero).
It is the outstanding mathematical problem to find the non-perturbative holomorphic prepotential
for such manifolds. Because of this problem, it might be reasonable to drop the assumption
of rigidness and consider more general CY threefolds. Since the D-instanton corrected metric
on the hypermultiplet moduli space is known for any CY \cite{Alexandrov:2014sya},
it may not be difficult to generalize the derivation of the non-perturbative potential \eqref{potential-main}
to a generic case. However, then both the metric and the scalar potential would become
even more complicated by acquiring extra dependence on the
complex structure moduli which also have to be stabilized.

Finally, it should be emphasized that we considered the very restricted set of fluxes, with all magnetic fluxes,
including Romans mass, being set to zero. It was chosen to preserve $N=2$ local supersymmetry that,
in turn, was needed to take into account non-perturbative contributions, which are known only under very special circumstances.
Of course, from both phenomenological and pure theoretical viewpoints,
it would be desirable to extend our analysis to more general flux compactifications when
$N=2$ local supersymmetry is broken to $N=1$. This, however, would require a much better understanding
of quantum effects in $N=1$ flux compactifications, beyond the current level.
Whereas their direct calculation from the first principles
is hardly possible, one may hope that
a combination of string dualities with geometry of the moduli spaces will become as powerful
in the $N=1$ case as it turned out to be in the $N=2$ case.

\acknowledgments

It is our pleasure to thank Sibasish Banerjee, Eric Bergshoeff, Renata Kallosh, Amir-Kian Kashani-Poor,
Ruben Minasian, Ulrich Theis and Stefan Vandoren for valuable discussions and correspondence.
We are particularly grateful to Eberhard Freitag for sharing with us his results about the intersection numbers
of a rigid Calabi-Yau manifold with Picard number two. SVK is also grateful to the University of Montpellier 
for kind hospitality extended to him during part of this investigation.

SVK was supported by a Grant-in-Aid of the Japanese Society for Promotion of Science (JSPS) under No.~26400252,
the World Premier International Research Centre Initiative (WPI Initiative),
MEXT, Japan, and the Competitiveness Enhancement Program of Tomsk Polytechnic University in Russia.

\appendix

\section{Conventions and normalizations}
\label{ap-norm}

\subsection{Special geometry relations}
\label{subsec-special}

A special K\"ahler manifold $\cMsk$ is determined by a holomorphic prepotential $F(X^I)$, a homogeneous function of degree 2.
The homogeneous coordinates $X^I$ are related to the coordinates on the manifold $z^i$ by $z^i=X^I/X^0$ and,
for simplicity, we choose the gauge where $X^0=1$. Given the prepotential, it is convenient to define the matrix
\be
N_{IJ}=-2\Im F_{IJ}.
\label{defN}
\ee
It is invertible, but has a split signature $(b_2,1)$. A related invertible matrix with a definite signature
can be constructed as follows. Let us define
\be
\cN_{IJ}=\bF_{IJ}-\frac{\I \, N_{IK}X^K N_{JL}X^L}{N_{MN}X^M X^N}\,.
\label{defcN}
\ee
$\cN_{IJ}$ appears as the coupling matrix of the gauge fields in the low-energy effective action
and its imaginary part is negative definite.

In terms of the matrix \eqref{defN}, the K\"ahler potential on $\cMsk$ is given by
\be
\cK=-\log\(X^I N_{IJ}\bX^J\).
\label{defcK}
\ee
For reader's convenience, we give here its derivatives with respect to $z^i$ and $\bz^{\bi}$,
\begin{subequations}
\bea
\cK_i&=&-e^{\cK} N_{i I}\bX^I,
\label{derK}\\
\cK_{i\bj} &=&-e^{\cK} N_{ij}+\cK_i\cK_{\bj},
\label{derKKb}\\
\cK_{ij}&=&
-2 e^{\cK} F_{ijk}t^k+\cK_i\cK_j,
\label{derKK}\\
\cK_{ij\bk}&=&-\I e^{\cK} F_{ijk}+\cK_i \cK_{j\bk}+\cK_j\cK_{i\bk}+\cK_{\bk}\cK_{ij}-\cK_i\cK_j\cK_{\bk},
\label{derKKK}
\eea
\end{subequations}
where we have used homogeneity of the holomorphic prepotential and $z^i=b^i+\I t^i$.
In particular, \eqref{derKKb} provides the metric on $\cMsk$.

The inverse matrices of $N_{IJ}$, $\Im\cN_{IJ}$ and $\cK_{i\bj}$ are explicitly given by
\begin{subequations}
\bea
N^{IJ}&=&\Delta^{-1}\(\begin{array}{cc}
1 & -\hN^{jk}N_{0k}
\\
-\hN^{ik}N_{0k} & \hN^{ij}\Delta +\hN^{ik} N_{0k}\hN^{jl} N_{0l}
\end{array}\),
\label{invN}
\\
\hf\,\Im\cN^{IJ}&=&N^{IJ}-e^\cK\(X^I\bX^J+\bX^I X^J\),
\label{relNN}
\\
\cK^{i\bj}&=&-e^{-\cK}\(\hN^{ij}+\frac{e^{-\cK}\hN^{ik}\cK_{\bk}\hN^{jl}\cK_l}{1-e^{-\cK}\hN^{kl}\cK_{\bk}\cK_l}\),
\label{invcK}
\eea
\end{subequations}
where $\hN^{ij}$ denotes the inverse of $N_{ij}$ and $\Delta=N_{00}-\hN^{ij}N_{0i}N_{0j}$.
It follows from \eqref{invcK} that
\begin{subequations}
\bea
\cK^{i\bj}\cK_{\bj}&=&\frac{e^{-\cK} \hN^{ij}\cK_{\bj}}{e^{-\cK}\hN^{m\bn}\cK_m\cK_{\bn}-1}\, ,
\label{relKK}
\\
\cK_i\cK^{i\bj}\cK_{\bj}-1&=&\frac{1}{e^{-\cK}\hN^{m\bn}\cK_m\cK_{\bn}-1}\, .
\label{relKKK}
\eea
\end{subequations}
Finally, using the covariant derivatives
\be
D_i X^I=\(\p_i+\p_i\cK\) X^I,
\label{defDX}
\ee
we can rewrite \eqref{invN} and \eqref{relNN} as
\begin{subequations}
\bea
N^{IJ}&=&-e^\cK\(\cK^{i\bj}D_i X^I D_{\bj}\bX^J-X^I\bX^J\),
\\
\hf\,\Im \cN^{IJ} &=& -e^\cK\(\cK^{i\bj}D_i X^I D_{\bj}\bX^J+\bX^I X^J\).
\label{relcND}
\eea
\end{subequations}

\subsection{Quaternionic geometry and scalar potential}
\label{subsec-qk-pot}

There is the extensive literature about the scalar potential in the gauged $N=2$ supergravity,
although one should be careful of the relative normalization of various contributions
to the scalar potential in explicit calculations. There are two sources of different normalizations:
(i) the gravitational coupling constant $\kappa^2$,
and (ii) the relation between the quaternionic 2-forms and the metric tensor on the hypermultiplet moduli space.
Here we explain this numerical ambiguity, provide the scalar potential with a generic choice of the parameters,
fix our conventions and compare them to the literature.

We recall that a QK space is a $4n$-real-dimensional manifold with a holonomy group
$Sp(n)\times SU(2)$ \cite{MR664330}. It is characterized by the existence of
a quaternionic structure encoded in a triplet of 2-forms $\vec \omega$.
Let us choose the local complex structure such that $\omega^+$ is
holomorphic.\footnote{We define the chiral basis as
$x^\pm=-\hf(x^1\mp\I x^2)$, so that $\vec x\cdot \vec y=x^3y^3+2x^+ y^-+2x^- y^+$.}
Denoting by $\pi^X$ a basis of (1,0)-forms in this complex structure, the metric is related to $\omega^3$ as
\be
\de s^2=2 g_{X\bY}\pi^X\otimes\bar\pi^Y,
\qquad
\omega^3=\I a \,g_{X\bY}\pi^X\otimes\bar\pi^Y,
\label{def-gom}
\ee
where we have parametrized the ambiguity in the normalization of the quaternionic forms by a real parameter $a$.

The triplet of 2-forms $\vec\omega$ is, in fact, proportional to the $SU(2)$ part of the curvature.
Denoting by $\vec p$ the $SU(2)$ part of the spin connection, we have
\be
\de \vec p+\hf\, \vec p\times \vec p=\frac{\nu}{a}\,\vec \omega,
\label{pom}
\ee
where $\nu$ is the constant of proportionality.
It is related to the scalar curvature and the quaternionic dimension $n$ as $R=4n(n+2)\nu$
and, hence, fixes the scale of the metric. When $n=1$, we find $R=12\nu$ and, consequently,
$\nu=\Lambda/3$, where $\Lambda$ is a ``cosmological constant".
Furthermore, given the conventional normalization of the kinetic terms of graviton and scalars,
\be
L_{\rm kin}=-\frac{1}{2\kappa^2}\, R(e)-\hf \, g_{uv}\p_\mu\phi^u\p^\mu\phi^v,
\ee
the local $N=2$ supersymmetry fixes the parameter $\nu$ in terms of the gravitational coupling as $\nu=-\kappa^2$.

Finally, when the QK metric has a Killing vector $k\in T\cM$, one can define a triplet of moment maps $\vec\mu$
\cite{MR872143} by the equation
\be
\p_u\vec \mu+\vec p_u\times \vec \mu=\vec \omega_{uv} k^v
\ \Rightarrow\ \vec \mu=-\frac{1}{2\nu}\, \vec\omega^u_{\ v}D_u k^v,
\label{defmu}
\ee
where we have used \eqref{pom} and the fact that $\vec\omega$ are covariantly constant with respect to the $SU(2)$ connection,
$\de\vec\omega+\vec p\times\vec\omega=0$.

Given these conventions with arbitrary $\kappa^2$ and $a$, the scalar potential resulting from gauging
a set of isometries on the hypermultiplet moduli space is given by
\be
\begin{split}
V =&\, 2\kappa^{-2} e^\cK \bfk^u_I \bfk^v_J g_{uv} X^I \bX^J -a^{-2}\(\hf\, \Im \cN^{IJ}+4e^\cK X^I \bX^J\)\(\vec\mu_I\cdot \vec \mu_J\)
\\
=&\, 2\kappa^{-2} e^\cK \bfk^u_I \bfk^v_J g_{uv} X^I \bX^J +a^{-2} e^\cK \(\cK^{i\bj}D_i X^I D_{\bj}\bX^J-3 X^I \bX^J\)\(\vec\mu_I\cdot \vec \mu_J\),
\end{split}
\label{scpot-generic}
\ee
where
$\bfk_I\in T\cM_H$ are the charge vectors characterizing the gauging and given by linear combinations of the
Killing vectors, and $\vec \mu_I$ are the triplets of moment maps defined by $\bfk_I$. In
going from the first representation to the second one, we have used the relation \eqref{relcND}.
Note that the definition \eqref{defmu} implies $\vec \mu\sim a\kappa^{-2}$, whereas a combination of \eqref{def-gom} and
\eqref{pom} yields $g_{uv}\sim \kappa^{-2}$.
As a result, using all these facts in the potential \eqref{scpot-generic}, we find that the dependence on
all normalization constants is factorized as $V\sim \kappa^{-4}$ so that different normalizations do not affect physics.

In the literature, one can find two natural choices for the parameter $a$ and two choices for $\kappa^2$:
\begin{center}
\begin{tabular}{c|c|c}
& $a=1$ & $a=2$
\\ \hline
$\vphantom{\frac{A^{A^A}}{A_{A_A}}}$
$\kappa^2=2$  & \cite{deWit:2001bk} & \cite{Fre:2014pca}
\\ \hline
$\vphantom{\frac{A^{A^A}}{A_{A_A}}}$
$\kappa^2=a/2$ & \cite{Davidse:2005ef,Alexandrov:2014sya} & \cite{Alexandrov:2011va}
\end{tabular}
\end{center}
Substituting the corresponding values of $\kappa^2$ and $a$, one can check that the scalar potential \eqref{scpot-generic}
agrees with the one given in \cite{Davidse:2005ef}, but differs by the factor of $1/4$ from
those in \cite{deWit:2001bk,Fre:2014pca}. In this paper, we choose
$a=1$ and $\kappa^2=1/2$ so that the scalar potential takes the form \eqref{scpot-gen}.

\section{The worldsheet instanton corrected metric on $\cM_V$}
\label{ap-VMmetric}

Here we apply the general special geometry relations given in Appendix \ref{subsec-special}
to the case of the holomorphic prepotential \eqref{Ffull} defining the geometry of
$N=2$ vector multiplets in the type IIA compactifications. A straightforward calculation yields
\begin{subequations}
\bea
N_{ij}&=&2\kappa_{ijk}t^k+\frac{1}{2\pi}\sum_{k_l\gamma^l\in H_2^+(\CY)}\nk k_i k_j\log\left|1-e^{2\pi\I k_l z^l}\right|^2,
\label{resNij}\\
N_{0i}&=&-2 \kappa_{ijk}b^j t^k
-\frac{1}{2\pi^2}\!\!\sum_{k_l\gamma^l\in H_2^+(\CY)}\!\!\nk k_i \Im  \[ \Li_2\(e^{2\pi\I k_l z^l}\)+2\pi\I k_jz^j\log\(1-e^{2\pi\I k_l z^l}\)\]
\\
N_{iI}\bX^I &=& -e^{-\cK}\cK_i=-2\I \kappa_{ijk}t^j t^k
\label{resNX}\\
&&
+\frac{\I}{4\pi^2}\sum_{k_l\gamma^l\in H_2^+(\CY)}\nk k_i \[\Li_2\(e^{2\pi\I k_l z^l}\)-\Li_2\(e^{-2\pi\I k_l \bz^l}\)
-4\pi k_j t^j \log\(1-e^{2\pi\I k_l z^l}\)\],
\nn\\
e^{-\cK} &=& 8\cV -C
+\frac{1}{2\pi^3}\sum_{k_l\gamma^l\in H_2^+(\CY)}\nk \Re\[\Li_3\(e^{2\pi\I k_l z^l}\)
+2\pi k_j t^j \Li_2\(e^{2\pi\I k_l z^l}\)\],
\label{resK}
\eea
\label{res-FVM}
\end{subequations}
where we have introduced the CY volume, $\cV=\frac16\, \kappa_{ijk}t^i t^j t^k$, and the parameter controlling
the perturbative $\alpha'$-correction
\be
C=\frac{\zeta(3)\chi_\CY}{4\pi^3}\, .
\label{defC}
\ee
The metric in question is obtained by plugging these results into \eqref{derKKb}.

The inverse metric and other related matrices cannot be explicitly computed in the presence of worldsheet instantons.
If, however, we restrict ourselves to the perturbative approximation,
then they can be expressed in terms of the inverse of $\kappa_{ij}\equiv\kappa_{ijk}t^k$ which we denote by $\kappa^{ij}$.
In particular, we find
\begin{subequations}
\bea
\cK_i & \approx & \frac{2\I\kappa_{ij} t^j}{8\cV-C}~~,
\\
\cK_{i\bj}&\approx & -\frac{2\kappa_{ij}}{8\cV-C}+\frac{4\kappa_{ik} t^k \kappa_{jl}t^l}{(8\cV-C)^2}~~
\\
\cK^{i\bj} &\approx & -\hf(8\cV-C)\(\kappa^{ij}-\frac{2t^it^j}{4\cV+C}\),
\\
N_{IJ}&\approx & \(\begin{array}{cc}
2\kappa_{ij}b^ib^j-4\cV-C\ & -2\kappa_{ij}b^j
\\
-2\kappa_{ij}b^j & 2\kappa_{ij}\end{array}\),
\\
N^{IJ}&\approx & -\frac{1}{4\cV+c}\(\begin{array}{cc}
1 & b^i
\\
b^i &\ b^ib^j-\hf\,(4\cV+C)\kappa^{ij}\end{array}\),
\\
-\hf\, \Im \cN^{IJ}&\approx & \(\frac{1}{4\cV+C} +\frac{2}{8\cV-C}\)\(\begin{array}{cc}
1 & b^i \\ b^i & b^i b^j \end{array}\)
+\(\begin{array}{cc}
0 & 0
\\
0 &\ \frac{2t^it^j}{8\cV-C}-\hf\,\kappa^{ij}\end{array}\).
\label{pert-cNinv}
\eea
\end{subequations}

\section{The D-instanton corrected metric of the universal hypermultiplet}
\label{ap-UHMmetric}

\subsection{The metric}
\label{subap-metric}

To write down the metric computed in \cite{Alexandrov:2014sya}, we have to introduce several important objects.
First, let us summarize the data characterizing a rigid Calabi-Yau manifold:
\begin{itemize}
\item
The intersection numbers $\kappa_{ijk}$, which specify the classical holomorphic prepotential \eqref{Fcl}
on the K\"ahler moduli space.

\item
The Euler characteristic $\chi_\CY=2h^{1,1}>0$, which appears in the $\alpha'$-corrected prepotential \eqref{Ffull},
and is always positive for rigid $\CY$. We also use the following parameter:
\be
c=-\frac{\chi_\CY}{192\pi}=-\frac{\pi^2}{48\zeta(3)}\, C.
\label{def-c}
\ee

\item
The complex number
\be
\lambda\equiv \lambda_1 -\I \lambda_2=\frac{\int_\cB\Omega}{\int_\cA\Omega}
\label{prepUHM}
\ee
given by the ratio of periods of the holomorphic 3-form $\Omega\in H^{3,0}(\CY)$
over an integral symplectic basis $(\cA,\cB)$ of $H_3(\CY,\IZ)$. The geometry requires that $\lambda_2>0$,
which explains the minus sign in \eqref{prepUHM}.

\item
The generalized Donaldson-Thomas (DT) invariants $\Omega_\gamma$,
which are integers counting, roughly, the number of BPS instantons
of charge $\gamma=(p,q)$. In the case of the vanishing magnetic charge $p$
and arbitrary electric charge $q$, they coincide with the Euler characteristic,
\be
\Omega_{(0,q)}=\chi_\CY.
\label{Omq}
\ee
\end{itemize}

Next, we introduce the central charge
\be
Z_\gamma=q-\lambda p
\label{cencharge}
\ee
which characterizes a D-instanton of charge $\gamma$.
It is used to define the function
\be
\cX_\gamma(t)=(-1)^{qp} \exp\[-2\pi\I\(q\zeta-p\tzeta+\cR\(\varpi^{-1}Z_\gamma-\varpi \bZ_\gamma\)\)\],
\label{defcX}
\ee
where $\cR$ is a function on the moduli space, which is fixed below.
Geometrically, $t$ parametrizes the fiber of the twistor space $\cZ$, a $\CP$ bundle over $\cM_H$,
whereas $\cX_\gamma$ are Fourier modes of holomorphic Darboux coordinates on
$\cZ$ \cite{Alexandrov:2008nk,Alexandrov:2008gh}. Using \eqref{defcX}, we define
\be
\begin{array}{rclrcl}
\Igg{}& = &\displaystyle
\int_{\ellg{\gamma}}\frac{\d \varpi}{\varpi}\,
\log\(1-\cX_\gamma\),
\qquad &
\rIg&=&
\displaystyle \int_{\ellg{\gamma}}\frac{\d \varpi}{\varpi}\,
\frac{\cX_\gamma}{1-\cX_\gamma}\, ,
\\
\Igpm& = &\displaystyle
\pm\int_{\ellg{\gamma}}\frac{\d \varpi}{\varpi^{1\pm 1}}\,\log\(1-\cX_\gamma\),
\qquad &
\rIgpm&=&
\displaystyle
\pm\int_{\ellg{\gamma}}\frac{\d \varpi}{\varpi^{1\pm 1}}\,
\frac{\cX_\gamma}{1-\cX_\gamma}\, ,
\end{array}
\label{def-JJJ}
\ee
where $\ellg{\gamma}$ is a contour on $\CP$ joining $t=0$ and $t=\infty$ along the direction fixed by the phase
of the central charge, $\ellg{\gamma}=\I Z_\gamma \IR^+$.
These functions satisfy the reality properties
\be
\overline{\Ingam{n}{\gamma}}=\Ingam{n}{-\gamma},
\qquad
\overline{\Insgam{n}{+}{\gamma}}=\Insgam{n}{-}{-\gamma},
\label{real-JJJ}
\ee
and the following identities:
\be
\Zg{}\Insgam{n}{+}{\gamma}=\bZg{}\Insgam{n}{-}{\gamma},
\label{ident-JJJ}
\ee
which can be verified by partial integration. Expanding the integrands in powers of $\cX_\gamma$,
they can be expressed as series of the modified Bessel functions of the second kind $K_n$.

The set of functions \eqref{def-JJJ} encodes the D-instanton corrections to the moduli space.
It is, however, convenient to introduce a few more quantities, which explicitly appear in the metric:
\begin{itemize}
\item
the functions
\be
\begin{split}
\vl =\frac{1}{2\pi}\sum_\gamma \Om{\gamma} |Z_\gamma|^2 \rIgm,
&\qquad
\Min= 2\lambda_2-\frac{1}{2\pi}\sum_\gamma \Om{\gamma} |Z_\gamma|^2 \rIg,
\\
\Uin =&\,\Min+\Min^{-1}|\vl|^2,
\end{split}
\label{Ab-UHM}
\ee

\item
the one-forms
\bea
\hspace{-1.2cm}
\cY &=& \de\tzeta-\lambda \de\zeta-\frac{\I}{4\pi}\sum_\gamma \Om{\gamma}
Z_\gamma \(\rIg-\vl\Min^{-1}\rIgp\)\(q\de \zeta-p\de\tzeta\)
-\frac{2\I\vl}{ \cR \Min}\, \de r,
\\
\hspace{-1.2cm}
\cVs &=&\frac{2r}{\pi\lambda_2\cR\Uin}\sum_\gamma \Om{\gamma} Z_\gamma
\(\rIgp+\bvl\Min^{-1} \rIg\)\[\(q-\lambda_1 p\)\(\de\tzeta-\lambda_1\de\zeta\)+\lambda_2^2 p\de\zeta \].
\label{conn-UHM}
\eea
\end{itemize}

Finally, the function $\cR$ entering \eqref{defcX} is implicitly determined as a solution to the following equation:
\be
r= \frac{\lambda_2\cR^2}{2}-c
-\frac{ \I\cR}{32\pi^2}\sum\limits_{\gamma}\hng{} \(Z_\gamma\Igp+\bZ_\gamma\Igm\),
\label{r-UHM}
\ee
where $r=e^\phi$ is the four-dimensional dilaton.

With all the notations above, the D-instanton corrected metric on the four-dimensional hypermultiplet moduli space is given by
\be
\de s^2=
\frac{2}{r^2}\[\(1-\frac{2r}{\cR^2\Uin}\) \((\de r)^2+\frac{\cR^2}{4}\,|\cY|^2\)
+\frac{1}{64}\(1-\frac{2r}{\cR^2\Uin}\)^{-1}\(\de \sigma +\tzeta \de \zeta-\zeta\de \tzeta+\cVs \)^2\].
\label{mett-UHM}
\ee
There are two regimes of validity of this result:
\begin{itemize}
\item
One includes D-instantons of all charges $\gamma=(p,q)$, but in this case the metric
is not valid beyond the one-instanton approximation, i.e. only the terms {\it linear}
in the DT invariants $\Omega_\gamma$ can be trusted. In particular,
such metric is merely {\it approximately} quaternion-K\"ahler.
In this approximation it can be further simplified by expanding all the coefficients to the first order in $\Omega_\gamma$.
\item
One includes only ``half" of D-instantons by restricting them to a set of charges satisfying the condition of mutual locality,
$\langle\gamma,\gamma'\rangle=0$. In this case, all charge vectors are proportional to each other,
$\gamma=n\gamma_0$ with a fixed $\gamma_0$. This restriction should be imposed
in all sums over $\gamma$ in \eqref{Ab-UHM}-\eqref{r-UHM}.
Then the metric is {\it exactly} quaternion-K\"ahler.
\end{itemize}

As was proven in \cite{Alexandrov:2014sya} in the latter case, the metric \eqref{mett-UHM}
agrees with the Tod ansatz \cite{MR1423177} where the role of the Tod potential satisfying the Toda equation
is played by the function $T=2\log(\cR/2)$. Unfortunately, this function and the coordinates adapted to the Tod ansatz
are defined only implicitly in terms of the physical fields, so that we do not use this representation here.

\subsection{The moment maps}
\label{subap-moment}

In this Appendix we evaluate the quaternionic moment maps for the Killing vectors \eqref{kilv-H} of the metric \eqref{mett-UHM}.
They are defined by equation \eqref{defmu} and
can be found using the following trick. Let us contract \eqref{pom} with $\bfk_I$.
In our normalization $\nu a^{-1}=-\hf$, so that the resulting equation can be written down as
\be
\iota_{\bfk_I}\vec \omega=2\de \(\iota_{\bfk_I}\vec p\)+2 \vec p\times\(\iota_{\bfk_I}\vec p\)-2\cL_{\bfk_I}\vec p\,.
\label{pomk}
\ee
Thus, if the following condition holds
\be
\cL_{\bfk_I}\vec p=\iota_{\bfk_I}\de\vec p+d\(\iota_{\bfk_I}\vec p\)=0,
\label{vanishLie}
\ee
comparing \eqref{pomk} and \eqref{defmu}, we conclude that the moment maps are given by
a very simple expression,
\be
\vec \mu_I=2\iota_{\bfk_I}\vec p\,.
\label{momentmap}
\ee

To proceed further, we have to know the explicit expressions of the components of the $SU(2)$ connection $\vec p$.
They were computed in \cite{Alexandrov:2014sya} and, after restricting them to four dimensions, the result reads
\be
\begin{split}
p^+ =&\,\frac{\I\cR}{4r}
\( \de\tzeta-\lambda\de\zeta-\frac{1}{8\pi^2}\sum\limits_{\gamma} \hng{} \Zg{}\de \Igg{}\) ,
\\
p^3 =&\, \frac{1}{8r} \( \de\sigma
+\tzeta\de \zeta-\zeta\de \tzeta\).
\end{split}
\label{SU(2)connect}
\ee

First, we need to check the vanishing of Lie derivatives \eqref{vanishLie} along the Killing vectors \eqref{kilv-H}.
It is a trivial exercise for $p^3$, whereas for $p^+$ we find
\be
\cL_{\bfk_I}p^+=\frac{\I}{4r}\( \iota_{\bfk_I}\de\cR\)\!
\( \de\tzeta-\lambda\de\zeta-\frac{1}{8\pi^2}\sum\limits_{\gamma} \hng{} \Zg{}\de \Igg{}\)
-\frac{\I\cR}{32\pi^2 r}\sum\limits_{\gamma} \hng{} \Zg{}\de\(\iota_{\bfk_I}\de \Igg{}\)\!.
\label{Liepplus}
\ee
Let us now take into account that the presence of the $H$-flux only admits D-instantons of the charges
satisfying $hq=\thh p$. This implies that $\iota_{\bfk_I}\de\cX_\gamma=0$ and, as a result,
both terms in \eqref{Liepplus} vanish. Due to the same reason, the instanton term in \eqref{SU(2)connect}
does not contribute to the moment maps \eqref{momentmap}.
Then it is easy to check that they coincide with the expressions given in \eqref{momentmap-main}.

\section{The second derivatives of the perturbative potential}
\label{ap-second}

A straightforward calculation gives the following results for the second derivatives of the scalar potential \eqref{pertpot}:
\begin{subequations}
\bea
\p_r^2 \Va &=& \frac{e^{\cK}}{4r^4}\[ \frac{16\thh^2}{\lambda_2}\,\frac{(1-\gp)r+6c}{1+\gp}
+\frac{4(et)^2 r}{(r+2c)^3}\(3(r+c)^2+c^2\)-3e^{-\cK}\kappa^{ij}e_i e_j\],
\\
\p_{t^i}\p_r \Va &=& \frac{e^{\cK}}{4r^3}\Biggl[
16 e^{\cK}\kappa_{ij}t^j\( \frac{2\thh^2}{\lambda_2}\,\(\frac{2(r+2c)}{(1+\gp)^2}-r\)+\frac{(et)^2 r(r+c)}{(r+2c)^2}\)
-\frac{8(et)e_i r(r+c)}{(r+2c)^2}
\Biggr.
\nonumber\\
&&\Biggl.\qquad
-e^{-\cK}\kappa_{ijk}\,\kappa^{jm} e_m \,\kappa^{kn}e_n \Biggr],
\\
\p_{t^i}\p_{t^j} \Va &=& \frac{e^{\cK}}{4r^2}\Biggl[
64 e^{2\cK}\kappa_{ik}t^k\kappa_{jl} t^l\( \frac{4\thh^2}{\lambda_2}\,\(\frac{2(r+c)}{(1+\gp)^3}-r\)+\frac{r(et)^2 }{r+2c}\)
\Biggr.
\nonumber\\
&&\qquad
-16 e^{\cK}\kappa_{ij}\( \frac{4\thh^2}{\lambda_2}\,\(\frac{2(r+c)}{(1+\gp)^2}-r\)+\frac{r(et)^2}{r+2c}\)
-\frac{16 r(et) e^{\cK}}{r+2c}\( e_i \kappa_{jk}+e_j\kappa_{ikl}\)t^k
\nonumber\\
&&\Biggl.\qquad
+\frac{4r e_i e_j}{r+2c}
-e^{-\cK}\kappa_{ikl}\,\kappa^{kp}e_p\,\kappa^{lm}\,\kappa_{jmn}\,\kappa^{nq} e_q \Biggr].
\eea
\end{subequations}
Contracting the last two quantities with $t^i$ and $n^i\equiv \frac{\kappa^{ij}e_j}{e^{\cK}(et)}$, we find
\begin{subequations}
\bea
t^i\p_{t^i}\p_r \Va &=&
\frac{e^{\cK}}{4r^3}\[ \frac{8\thh^2(3+\gp)}{\lambda_2}\(\frac{2(r+2c)}{(1+\gp)^2}-r\)
+\frac{4(et)^2 r(r+c)(1+\gp)}{(r+2c)^2}-e^{-\cK}\kappa^{ij}e_i e_j\],
\\
n^i\p_{t^i}\p_r \Va &=&
\frac{e^{\cK}}{4r^3}\Biggl[ \frac{32\thh^2}{\lambda_2}\(\frac{2(r+2c)}{(1+\gp)^2}-r\)
+\frac{8 r(r+c)}{(r+2c)^2}\(2(et)^2-e^{-\cK}\kappa^{ij}e_i e_j\)
\Biggr.
\nonumber\\
&&\Biggl. \qquad
-\frac{e^{-2\cK}}{(et)}\,\kappa_{ijk}\,\kappa^{il}e_l\,\kappa^{jm} e_m\, \kappa^{kn}e_n\Biggr],
\eea
\bea
t^it^j\p_{t^i}\p_{t^j} \Va &=&
\frac{e^{\cK}}{4r^2}\Biggl[ \frac{16\thh^2(3+\gp)}{\lambda_2}\(\frac{4(r+c)}{(1+\gp)^3}-(2+\gp)r\)
+\frac{4(et)^2 r(1+3\gp+\gp^2)}{r+2c}
\Biggr.
\nonumber\\
&&\Biggl. \qquad
-e^{-\cK}\kappa^{ij}e_i e_j\Biggr],
\\
n^i t^j\p_{t^i}\p_{t^j} \Va &=&
\frac{e^{\cK}}{4r^2}\Biggl[ \frac{64\thh^2}{\lambda_2}\(\frac{4(r+c)}{(1+\gp)^3}-(2+\gp)r\)
+\frac{4r}{r+2c}\(4(1+\gp)(et)^2-e^{-\cK}(2+\gp)\kappa^{ij}e_i e_j\)
\nonumber\\
&&\Biggl.\qquad
-\frac{e^{-2\cK}}{(et)}\,\kappa_{ijk}\,\kappa^{il}e_l\,\kappa^{jm} e_m\, \kappa^{kn}e_n\Biggr],
\\
n^i n^j\p_{t^i}\p_{t^j} \Va &=&
\frac{e^{\cK}}{4 r^2(et)^2}\Biggl[ 16\(4(et)^2-e^{-\cK}\kappa^{ij}e_i e_j\)
\(\frac{4\thh^2}{\lambda_2}\(\frac{2(r+c)}{(1+\gp)^2}-r\)+\frac{(et)^2 r}{r+2c}\)
\Biggr.
\nonumber\\
&&\qquad
-\frac{512\thh^2(et)^2 \gp(r+c)}{\lambda_2(1+\gp)^3}
-\frac{4r\,e^{-\cK} }{r+2c}\,\kappa^{ij}e_i e_j \(8(et)^2-e^{-\cK}\kappa^{ij}e_i e_j\)
\nonumber\\
&&\Biggl. \qquad
-e^{-3\cK}\kappa^{ij}\kappa_{ikl}\,\kappa^{kp}e_p\,\kappa^{lq} e_q\, \kappa_{jmn}\,\kappa^{mr}e_r\, \kappa^{ns}e_s\Biggr].
\eea
\end{subequations}

Using \eqref{proj-stabKt}, \eqref{eq-et} and \eqref{proj-stabKn},
all terms with the intersection numbers and/or flux parameters $e_i$, and,
hence, all the second derivatives also, can be written down as functions of $\gp$ and $r$ only.
Explicitly, they are
\begin{subequations}
\bea
\p_r^2 \Va
&=& -\frac{2\thh^2 e^{\cK}}{\lambda_2 r^4}\,
\frac{\gp(1-\gp^2) r^3-2 (7 - 17 \gp - 13 \gp^2 - \gp^3)c r^2-8(4-3\gp-2\gp^2)c^2 r-16 c^3 }{(1+\gp)^2(r+2c)(\gp r+c(1+2\gp))},
\eea
\bea
t^i\p_{t^i}\p_r \Va
&=&
-\frac{2\thh^2 e^{\cK}}{\lambda_2 r^3}\, \frac{\gp(3-5\gp-5\gp^2-\gp^3)r^2-2(1-3\gp-2\gp^2)c r-8(1+\gp)c^2}{(1+\gp)^2(\gp r+c(1+2\gp))},
\\
n^i\p_{t^i}\p_r \Va
&=& -\frac{8\thh^2 e^{\cK}}{\lambda_2 r^3(1+\gp)^2(r+2c)^2 (\gp r+c(1+2\gp))}\,
\(\gp(1-2\gp-\gp^2)r^4
\right.
\\
&&\left. \qquad
+2(3-4\gp-11\gp^2-3\gp^3)c r^3+4\gp(3-4\gp-\gp^2)c^2r^2-8(5-2\gp)c^3 r-32 c^4\),
\nonumber
\eea
\bea
t^it^j\p_{t^i}\p_{t^j} \Va
&=&
-\frac{2\thh^2 e^{\cK}}{\lambda_2 r^2(1+\gp)^3(\gp r+c(1+2\gp))}
\( \gp(4+29\gp+5\gp^2-5\gp^3-\gp^4)r^2
\right.\\
&&\left.\qquad
-2(3-23\gp-69\gp^2-29\gp^3-2\gp^4)cr-4(3-4\gp-17\gp^2-6\gp^3)c^2\),
\nonumber
\\
n^i t^j\p_{t^i}\p_{t^j} \Va
&=&
\frac{8\thh^2 e^{\cK}}{\lambda_2 r^2(1+\gp)^3(r+2c)(\gp r+c(1+2\gp))}
\( \gp(2-13\gp-13\gp^2-7\gp^3-\gp^4)r^3
\right.
\nonumber\\
&&\left.\qquad
+2(1-8\gp-28\gp^2-4\gp^3-\gp^4)cr^2+4(2-13\gp-27\gp^2-2\gp^3)c^2r
\right.\\
&&\left.\qquad
+8(1-3\gp-6\gp^2)c^3\),
\nonumber
\\
n^i n^j\p_{t^i}\p_{t^j} \Va
&=&
\frac{32\thh^2 e^{\cK}}{\lambda_2 r^2(1+\gp)^3(r+2c)^3(\gp r+c(1+2\gp))((5-10\gp-9\gp^2- 2\gp^3) r+8c)}
\nonumber\\
&&\qquad \times
\( (100 - 320 \gp - 377 \gp^2 + 831 \gp^3 + 1246 \gp^4 + 614 \gp^5 + 135 \gp^6 + 11 \gp^7) r^6
\right.
\nonumber\\
&&\qquad
+2 (275 - 643 \gp - 1180 \gp^2 + 1232 \gp^3 + 2193 \gp^4 + 1043 \gp^5 + 216 \gp^6 + 16 \gp^7) c r^5
\nonumber\\
&&\qquad
+4 (168 - 233 \gp - 903 \gp^2 + 482 \gp^3 + 1087 \gp^4 + 439 \gp^5 + 76 \gp^6 + 4 \gp^7) c^2 r^4
\nonumber\\
&&\qquad
+8 (-158 + 139 \gp + 127 \gp^2 + 340 \gp^3 + 313 \gp^4 + 61 \gp^5 +6 \gp^6) c^3 r^3
\nonumber\\
&&\qquad
+16 (-232 + 21 \gp + 232 \gp^2 + 195 \gp^3 + 88 \gp^4 + 4 \gp^5) c^4 r^2
\nonumber\\
&&\left.\qquad
+ 32 (-93 - 36 \gp + 41 \gp^2 + 32 \gp^3 + 12 \gp^4) c^5 r-256 (3 + 2 \gp) c^6\).
\eea
\label{2der-final}
\end{subequations}

We are interested in whether the first three principle minors $\Delta_k$ of the matrix $\Mb$ \eqref{matder},
which are constructed from the quantities \eqref{2der-final},
can be simultaneously positive. Verifying this condition on Mathematica shows that
it is {\it never}\; satisfied (we employed the function RegionPlot with the argument
$\bigl\{\Delta_1>0\ \&\&\ \Delta_2>0\ \&\&\ \Delta_3>0\bigr\}$).
This implies that there are {\it no} local minima of the perturbative potential with at least two K\"ahler moduli.

\twofigdif{The functions $\Delta_1(\gp,r=r_+(\gp))$ (blue) and $\Delta_2(\gp,r=r_+(\gp))$ (red)
rescaled by $a\equiv\frac{3\lambda_2|c|C}{\thh^2}$ and $a^2$, respectively.
($\Delta_2$ was also multiplied by 10 to make it more visible.) The vertical line indicates the bound $\gp<\gp_\star$
in (4.16). The right picture magnifies the region near the critical value.
Both pictures demonstrate that, in the allowed range of $\gp$, the two functions are never simultaneously positive.}
{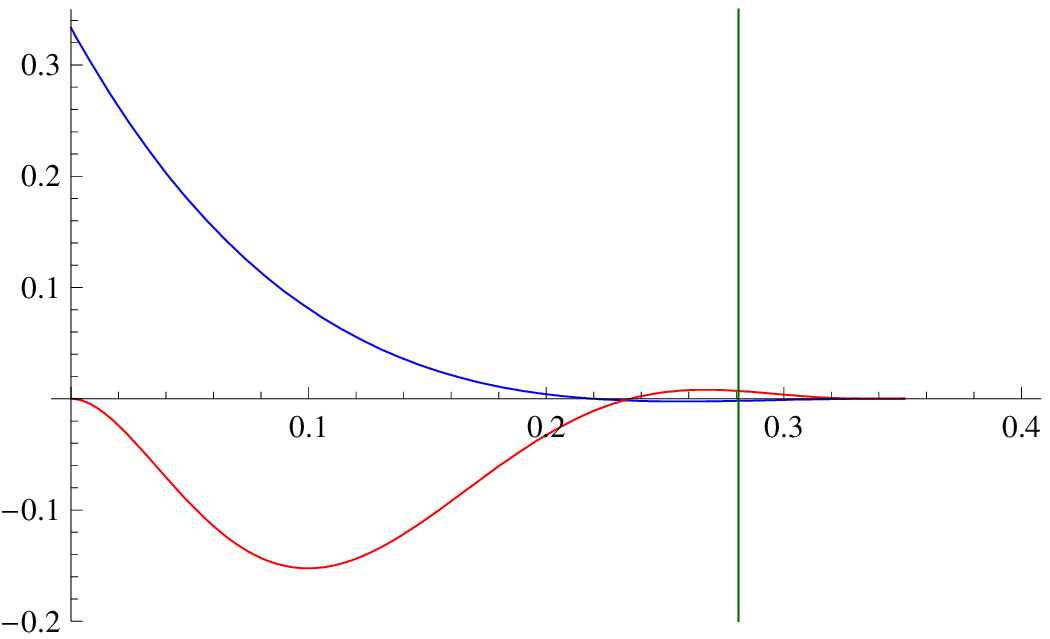}{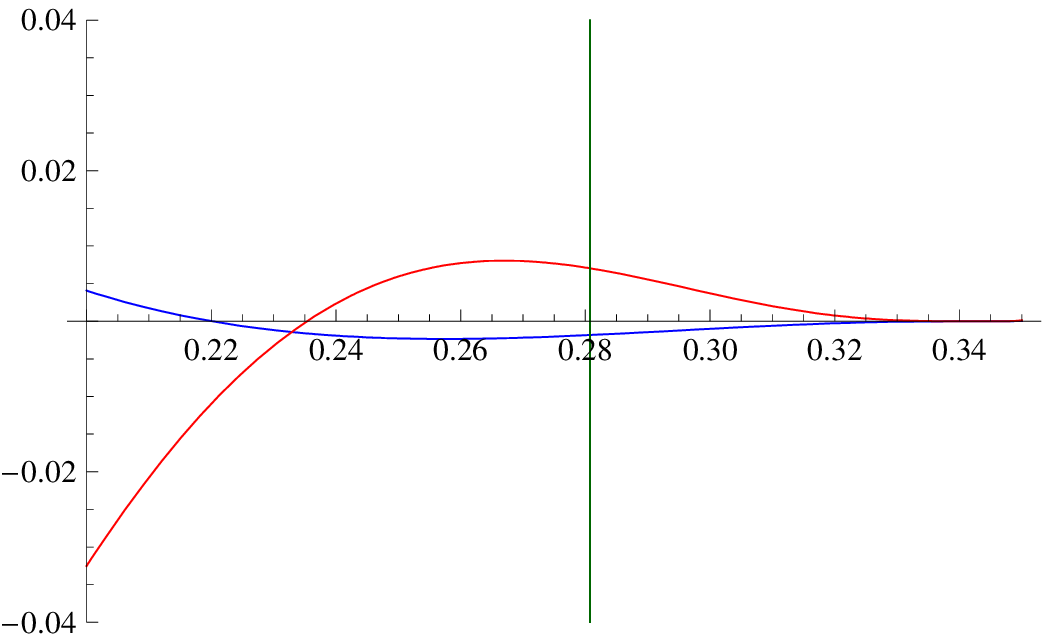}{9cm}{fig-2der-1mod}{0.4cm}{0.0cm}{7cm}

The same method can be used to show that the same conclusion remains true in the one-modulus case.
In this case, one should consider only two functions, $\Delta_1$ and $\Delta_2$.
Plotting the regions where they are positive shows that they have a non-trivial intersection.
Hence, this simple analysis does not exclude meta-stable vacua yet. However, in the one-modulus case
we know that critical points must belong to the curve $r=r_+(\gp)$ \eqref{roots}.
Substituting this into $\Delta_1$ and $\Delta_2$, and plotting the resulting functions of $\gp$,
we arrive at Fig. \ref{fig-2der-1mod}. It shows that these two functions are {\it never} simultaneously positive and,
hence, both critical points of the perturbative potential are unstable.

\section{The matrix $\Im\cN_{IJ}$}
\label{ap-cN}

In this Appendix we elaborate the condition of negative definiteness of the matrix $\Im\cN_{IJ}$
defined in \eqref{defcN}. Since any change of basis does not change the signature of a matrix,
we can equivalently consider the matrix $\mN=U^T (\Im\cN)\, U$.
Choosing $U=\(\begin{array}{cc} 1 & 0\\ b^i & {\delta^i}_j \end{array}\)$ and using
\be
N_{iI}X^I=-e^{-\cK}\cK_{\bi},
\qquad
N_{IJ}X^I X^J=e^{-\cK}\(1-2\I t^i\cK_{\bi}\),
\ee
we find
\be
\mN=\(\begin{array}{cc}
\hf\(e^{-\cK}-N_{ij}t^it^j\)-e^{-\cK}\Re\frac{\(1-\I t^i\cK_{\bi}\)^2}{1-2\I t^i\cK_{\bi}}
&\ -\hf \Re\cK_{\bi}+e^{-\cK}\Re\frac{\(1-\I t^i\cK_{\bi}\)\cK_{\bi}}{1-2\I t^i\cK_{\bi}}
\\
-\hf \Re\cK_{\bi}+e^{-\cK}\Re\frac{\(1-\I t^i\cK_{\bi}\)\cK_{\bi}}{1-2\I t^i\cK_{\bi}}
&
\hf\, N_{ij}- e^{-\cK}\Re\frac{\cK_{\bi}\cK_{\bj}}{1-2\I t^i\cK_{\bi}}
\end{array}\).
\label{matN}
\ee

Let us now take half-integer $b^i$-moduli as in \eqref{vanishsol}.
Then $\cK_{\bi}=-\I e^{\cK}N_{ij} t^j$ (see \eqref{res-FVM}) and the matrix \eqref{matN} simplifies to
\be
\mN=\(\begin{array}{cc}
-\hf\, e^{-\cK}\,\frac{e^{-\cK}-N_{ij}t^i t^j}{e^{-\cK}-2N_{ij}t^i t^j}
& 0
\\
0 &
\hf\, N_{ij}+ \frac{N_{ik}t^k N_{jl}t^l}{e^{-\cK}-2N_{ij}t^i t^j}
\end{array}\).
\label{matNsimple}
\ee
Thus, the negative definiteness of $\Im\cN_{IJ}$ requires
\be
\frac{N_{ij}t^i t^j-e^{-\cK}}{2N_{ij}t^i t^j-e^{-\cK}}>0
\qquad {\rm and}\qquad
\frac{2N_{ik}t^k N_{jl}t^l}{2N_{ij}t^i t^j-e^{-\cK}}-N_{ij}~~ \mbox{ is positive definite}.
\label{condmN}
\ee

Let us further restrict ourselves to the one-modulus case and drop the indices $i,j$ taking a single value.
Then the two conditions \eqref{condmN} simplify as
\be
\frac{N t^2-e^{-\cK}}{2Nt^2-e^{-\cK}}>0
\qquad {\rm and}\qquad
\frac{N e^{-\cK}}{2Nt^2-e^{-\cK}}>0.
\label{condNone}
\ee
Given positivity of $e^{-\cK}$, it is easy to see that these conditions are equivalent to \eqref{condNt}.

\providecommand{\href}[2]{#2}\begingroup\raggedright\endgroup

%\bibliographystyle{utphys}
%\bibliography{combined}

\end{document}